\pdfoutput=1
\documentclass[12pt,a4paper]{article}

\usepackage{ifthen} 
\newboolean{pdflatex}
\setboolean{pdflatex}{true} 

\newboolean{articletitles}
\setboolean{articletitles}{true} 

\newboolean{uprightparticles}
\setboolean{uprightparticles}{false} 

\newboolean{inbibliography}
\setboolean{inbibliography}{false} 

\usepackage{microtype}
\usepackage{lineno}  
\usepackage{xspace} 
\usepackage{caption} 

\usepackage{graphicx}  
\usepackage{color}
\usepackage{colortbl}
\graphicspath{{./figs/},{./figs-S/}} 

\usepackage{amsmath} 
\usepackage{amssymb}
\usepackage{amsfonts}
\usepackage{upgreek} 

\newcommand*\patchAmsMathEnvironmentForLineno[1]{%
\expandafter\let\csname old#1\expandafter\endcsname\csname #1\endcsname
\expandafter\let\csname oldend#1\expandafter\endcsname\csname
end#1\endcsname
 \renewenvironment{#1}%
   {\linenomath\csname old#1\endcsname}%
   {\csname oldend#1\endcsname\endlinenomath}%
}
\newcommand*\patchBothAmsMathEnvironmentsForLineno[1]{%
  \patchAmsMathEnvironmentForLineno{#1}%
  \patchAmsMathEnvironmentForLineno{#1*}%
}
\AtBeginDocument{%
\patchBothAmsMathEnvironmentsForLineno{equation}%
\patchBothAmsMathEnvironmentsForLineno{align}%
\patchBothAmsMathEnvironmentsForLineno{flalign}%
\patchBothAmsMathEnvironmentsForLineno{alignat}%
\patchBothAmsMathEnvironmentsForLineno{gather}%
\patchBothAmsMathEnvironmentsForLineno{multline}%
\patchBothAmsMathEnvironmentsForLineno{eqnarray}%
}

\usepackage{hyperref}    
\usepackage[all]{hypcap} 

\def\lhcb {\mbox{LHCb}\xspace}
\def\atlas  {\mbox{ATLAS}\xspace}
\def\cms    {\mbox{CMS}\xspace}

\def\lhc    {\mbox{LHC}\xspace}

\def\MagUp {\mbox{\em Mag\kern -0.05em Up}\xspace}

\ifthenelse{\boolean{uprightparticles}}%
{

 \def\Pmu         {\ensuremath{\upmu}\xspace}                 
 \def\Pnu         {\ensuremath{\upnu}\xspace}

 \def\Ptau        {\ensuremath{\uptau}\xspace}

 \def\PDelta      {\ensuremath{\Delta}\xspace}                 
 \def\PXi      {\ensuremath{\Xi}\xspace}                 
 \def\PLambda      {\ensuremath{\Lambda}\xspace}                 
 \def\PSigma      {\ensuremath{\Sigma}\xspace}                 
 \def\POmega      {\ensuremath{\Omega}\xspace}                 
 \def\PUpsilon      {\ensuremath{\Upsilon}\xspace}                 
 
 \def\PB      {\ensuremath{\mathrm{B}}\xspace}                 
                  
 \def\PD      {\ensuremath{\mathrm{D}}\xspace}

 \def\PK      {\ensuremath{\mathrm{K}}\xspace}

 \def\PW      {\ensuremath{\mathrm{W}}\xspace}

 \def\PZ      {\ensuremath{\mathrm{Z}}\xspace}                 
                  
 \def\Pb      {\ensuremath{\mathrm{b}}\xspace}                 
 \def\Pc      {\ensuremath{\mathrm{c}}\xspace}

 \def\Pi      {\ensuremath{\mathrm{i}}\xspace}

 \def\Pp      {\ensuremath{\mathrm{p}}\xspace}

 \def\Pt      {\ensuremath{\mathrm{t}}\xspace}

}
{

 \def\Pmu         {\ensuremath{\mu}\xspace}                 
 \def\Pnu         {\ensuremath{\nu}\xspace}

 \def\Ptau        {\ensuremath{\tau}\xspace}

 \mathchardef\PDelta="7101
 \mathchardef\PXi="7104
 \mathchardef\PLambda="7103
 \mathchardef\PSigma="7106
 \mathchardef\POmega="710A
 \mathchardef\PUpsilon="7107
                  
 \def\PB      {\ensuremath{B}\xspace}                 
                  
 \def\PD      {\ensuremath{D}\xspace}

 \def\PK      {\ensuremath{K}\xspace}

 \def\PW      {\ensuremath{W}\xspace}

 \def\PZ      {\ensuremath{Z}\xspace}                 
                  
 \def\Pb      {\ensuremath{b}\xspace}                 
 \def\Pc      {\ensuremath{c}\xspace}

 \def\Pi      {\ensuremath{i}\xspace}

 \def\Pp      {\ensuremath{p}\xspace}

 \def\Pt      {\ensuremath{t}\xspace}

}

\makeatletter
\ifcase \@ptsize \relax
  \newcommand{\miniscule}{\@setfontsize\miniscule{4}{5}}
\or
  \newcommand{\miniscule}{\@setfontsize\miniscule{5}{6}}
\or
  \newcommand{\miniscule}{\@setfontsize\miniscule{5}{6}}
\fi
\makeatother

\DeclareRobustCommand{\optbar}[1]{\shortstack{{\miniscule (\rule[.5ex]{1.25em}{.18mm})}
  \\ [-.7ex] $#1$}}

\def\mup        {{\ensuremath{\Pmu^+}}\xspace}
\def\mun        {{\ensuremath{\Pmu^-}}\xspace} 

\def\taup       {{\ensuremath{\Ptau^+}}\xspace}
\def\taum       {{\ensuremath{\Ptau^-}}\xspace}

\def\ellm       {{\ensuremath{\ell^-}}\xspace}
\def\ellp       {{\ensuremath{\ell^+}}\xspace}

\def\neu        {{\ensuremath{\Pnu}}\xspace}
\def\neub       {{\ensuremath{\overline{\Pnu}}}\xspace}

\def\W      {{\ensuremath{\PW}}\xspace}
\def\Wp     {{\ensuremath{\PW^+}}\xspace}
\def\Wm     {{\ensuremath{\PW^-}}\xspace}
\def\Wpm    {{\ensuremath{\PW^\pm}}\xspace}
\def\Z      {{\ensuremath{\PZ}}\xspace}

\def\cquark    {{\ensuremath{\Pc}}\xspace}

\def\bquark    {{\ensuremath{\Pb}}\xspace}

\def\tquark    {{\ensuremath{\Pt}}\xspace}
\def\tquarkbar {{\ensuremath{\overline \tquark}}\xspace}
\def\ttbar     {{\ensuremath{\tquark\tquarkbar}}\xspace}

  \def\Kbar    {{\kern 0.2em\overline{\kern -0.2em \PK}{}}\xspace}

\def\KorKbar    {\kern 0.18em\optbar{\kern -0.18em K}{}\xspace}

  \def\Dbar    {{\kern 0.2em\overline{\kern -0.2em \PD}{}}\xspace}

\def\DorDbar    {\kern 0.18em\optbar{\kern -0.18em D}{}\xspace}

\def\Bbar    {{\ensuremath{\kern 0.18em\overline{\kern -0.18em \PB}{}}}\xspace}

\def\BorBbar    {\kern 0.18em\optbar{\kern -0.18em B}{}\xspace}

  \def\Y#1S{\ensuremath{\PUpsilon{(#1S)}}\xspace}

\def\proton      {{\ensuremath{\Pp}}\xspace}

\def\Lbar        {{\ensuremath{\kern 0.1em\overline{\kern -0.1em\PLambda}}}\xspace}
\def\LorLbar    {\kern 0.18em\optbar{\kern -0.18em \PLambda}{}\xspace}

\newcommand{\decay}[2]{\ensuremath{#1\!\to #2}\xspace}         

\def\to                 {\ensuremath{\rightarrow}\xspace}

\newcommand{\as}{{\ensuremath{\alpha_s}}\xspace}

\def\AT#1     {\ensuremath{A_{\mathrm{T}}^{#1}}\xspace}           

\def\C#1      {\ensuremath{\mathcal{C}_{#1}}\xspace}                       
\def\Cp#1     {\ensuremath{\mathcal{C}_{#1}^{'}}\xspace}                    
\def\Ceff#1   {\ensuremath{\mathcal{C}_{#1}^{\mathrm{(eff)}}}\xspace}        
\def\Cpeff#1  {\ensuremath{\mathcal{C}_{#1}^{'\mathrm{(eff)}}}\xspace}       
\def\Ope#1    {\ensuremath{\mathcal{O}_{#1}}\xspace}                       
\def\Opep#1   {\ensuremath{\mathcal{O}_{#1}^{'}}\xspace}                    

\newcommand{\tev}{\ifthenelse{\boolean{inbibliography}}{\ensuremath{~T\kern -0.05em eV}\xspace}{\ensuremath{\mathrm{\,Te\kern -0.1em V}}}\xspace}
\newcommand{\gev}{\ensuremath{\mathrm{\,Ge\kern -0.1em V}}\xspace}
\newcommand{\mev}{\ensuremath{\mathrm{\,Me\kern -0.1em V}}\xspace}
\newcommand{\kev}{\ensuremath{\mathrm{\,ke\kern -0.1em V}}\xspace}
\newcommand{\ev}{\ensuremath{\mathrm{\,e\kern -0.1em V}}\xspace}
\newcommand{\gevc}{\ensuremath{{\mathrm{\,Ge\kern -0.1em V\!/}c}}\xspace}
\newcommand{\mevc}{\ensuremath{{\mathrm{\,Me\kern -0.1em V\!/}c}}\xspace}
\newcommand{\gevcc}{\ensuremath{{\mathrm{\,Ge\kern -0.1em V\!/}c^2}}\xspace}
\newcommand{\gevgevcccc}{\ensuremath{{\mathrm{\,Ge\kern -0.1em V^2\!/}c^4}}\xspace}
\newcommand{\mevcc}{\ensuremath{{\mathrm{\,Me\kern -0.1em V\!/}c^2}}\xspace}

\def\mum  {\ensuremath{{\,\upmu\rm m}}\xspace}

\def\pb {\ensuremath{\rm \,pb}\xspace}
\def\invpb {\ensuremath{\mbox{\,pb}^{-1}}\xspace}

\def\invfb   {\ensuremath{\mbox{\,fb}^{-1}}\xspace}

\def\gsim{{~\raise.15em\hbox{$>$}\kern-.85em
          \lower.35em\hbox{$\sim$}~}\xspace}
\def\lsim{{~\raise.15em\hbox{$<$}\kern-.85em
          \lower.35em\hbox{$\sim$}~}\xspace}

\def\sqs   {\ensuremath{\protect\sqrt{s}}\xspace}

\def\ptot       {\mbox{$p$}\xspace}
\def\pt         {\mbox{$p_{\rm T}$}\xspace}

\newcommand{\lum} {\ensuremath{\mathcal{L}}\xspace}

\def\evtgen     {\mbox{\textsc{EvtGen}}\xspace}
\def\fewz       {\mbox{\textsc{Fewz}}\xspace}
\def\dynnlo       {\mbox{\textsc{DYNNLO}}\xspace}

\def\geant      {\mbox{\textsc{Geant4}}\xspace}

\def\mc@nlo     {\mbox{\textsc{MC@NLO}}\xspace}

\def\photos     {\mbox{\textsc{Photos}}\xspace}
\def\powheg     {\mbox{\textsc{Powheg}}\xspace}
\def\pythia     {\mbox{\textsc{Pythia}}\xspace}
\def\resbos     {\mbox{\textsc{ResBos}}\xspace}

\def\tell1  {TELL1\xspace}
\def\ukl1   {UKL1\xspace}

\newcommand{\eg}{\mbox{\itshape e.g.}\xspace}

\usepackage{cite} 
\usepackage{mciteplus}

\usepackage{multirow}
\usepackage{bm}
\usepackage{rotating} 
\newcommand{\gec}{\mbox{\textsc{GEC}}\xspace}

\newcommand{\wmn}{\decay{\W}{\mu\neu}}
\newcommand{\wpmn}{\decay{\Wp}{\mup\neu}}
\newcommand{\wmmn}{\decay{\Wm}{\mun\neub}}

\newcommand{\wtn}{\decay{\W}{\tau\neu}}
\newcommand{\zmm}{\decay{\Z\,}{\mup\mun}}
\newcommand{\ztt}{\decay{\Z\,}{\taup\taum}}

\newcommand{\pp}{\ensuremath{\proton\proton}\xspace}

\newcommand{\csw}{\ensuremath{\sigma_{\wmn}}\xspace}
\newcommand{\cswp}{\ensuremath{\sigma_{\wpmn}}\xspace}
\newcommand{\cswm}{\ensuremath{\sigma_{\wmmn}}\xspace}

\newcommand{\csz}{\ensuremath{\sigma_{\zmm}}\xspace}

\newcommand{\ratiow}{\ensuremath{R_{\Wpm}}\xspace}
\newcommand{\ratiowz}{\ensuremath{R_{\W\Z}}\xspace}
\newcommand{\ratiowpz}{\ensuremath{R_{\Wp\Z}}\xspace}
\newcommand{\ratiowmz}{\ensuremath{R_{\Wm\Z}}\xspace}

\newcommand{\asy}{\ensuremath{A_{\mu}}\xspace}

\newcommand{\pdfs}{\mbox{\textsc{PDFs}}\xspace}

\newcommand{\nnlo}{\mbox{\textsc{NNLO}}\xspace}

\newcommand{\fsr}{\mbox{\textsc{FSR}}\xspace}

\newcommand{\rone}{\mbox{Run 1}\xspace}

\newcommand{\ptcone}{\ensuremath{p_{\textrm{T}}^{\textrm{cone}}}\xspace}
\newcommand{\etcone}{\ensuremath{E_{\textrm{T}}^{\textrm{cone}}}\xspace}
\newcommand{\etamu}{\ensuremath{\eta^{\mu}}\xspace}
\newcommand{\etamup}{\ensuremath{\eta^{\mu^{+}}}\xspace}
\newcommand{\etamum}{\ensuremath{\eta^{\mu^{-}}}\xspace}

\newcommand{\ip}{\textrm{IP}\xspace}

\newcommand{\eop}{\ensuremath{E_{\textrm{calo}} / \ptot c}\xspace}

\newcommand{\mmm}{\ensuremath{M_{\mu\mu}}\xspace}
\newcommand{\ptz}{\ensuremath{p_{\textrm{T},\Z}}\xspace}
\newcommand{\rapz}{\ensuremath{y_{\Z}}\xspace}

\newcommand{\phist}{\ensuremath{\phi^{*}_{\eta}}\xspace}

\newcommand{\effgec}{\ensuremath{\varepsilon_{\textrm{\gec}}}\xspace}
\newcommand{\effsel}{\ensuremath{\varepsilon_{\textrm{sel}}}\xspace}
\newcommand{\ffsr}{\ensuremath{f_{\textrm{\fsr}}}\xspace}

\newcommand{\herwigpp}{\mbox{\textsc{Herwig++}}\xspace}

\newcommand{\roounfold}{\mbox{\textsc{RooUnfold}}\xspace}

\usepackage{longtable} 

\begin{document}

\renewcommand{\thefootnote}{\fnsymbol{footnote}}
\setcounter{footnote}{1}

\begin{titlepage}
\pagenumbering{roman}

\vspace*{-1.5cm}
\centerline{\large EUROPEAN ORGANIZATION FOR NUCLEAR RESEARCH (CERN)}
\vspace*{1.5cm}
\noindent
\begin{tabular*}{\linewidth}{lc@{\extracolsep{\fill}}r@{\extracolsep{0pt}}}
\ifthenelse{\boolean{pdflatex}}
{\vspace*{-3.1cm}\mbox{\!\!\!\includegraphics[width=.14\textwidth]{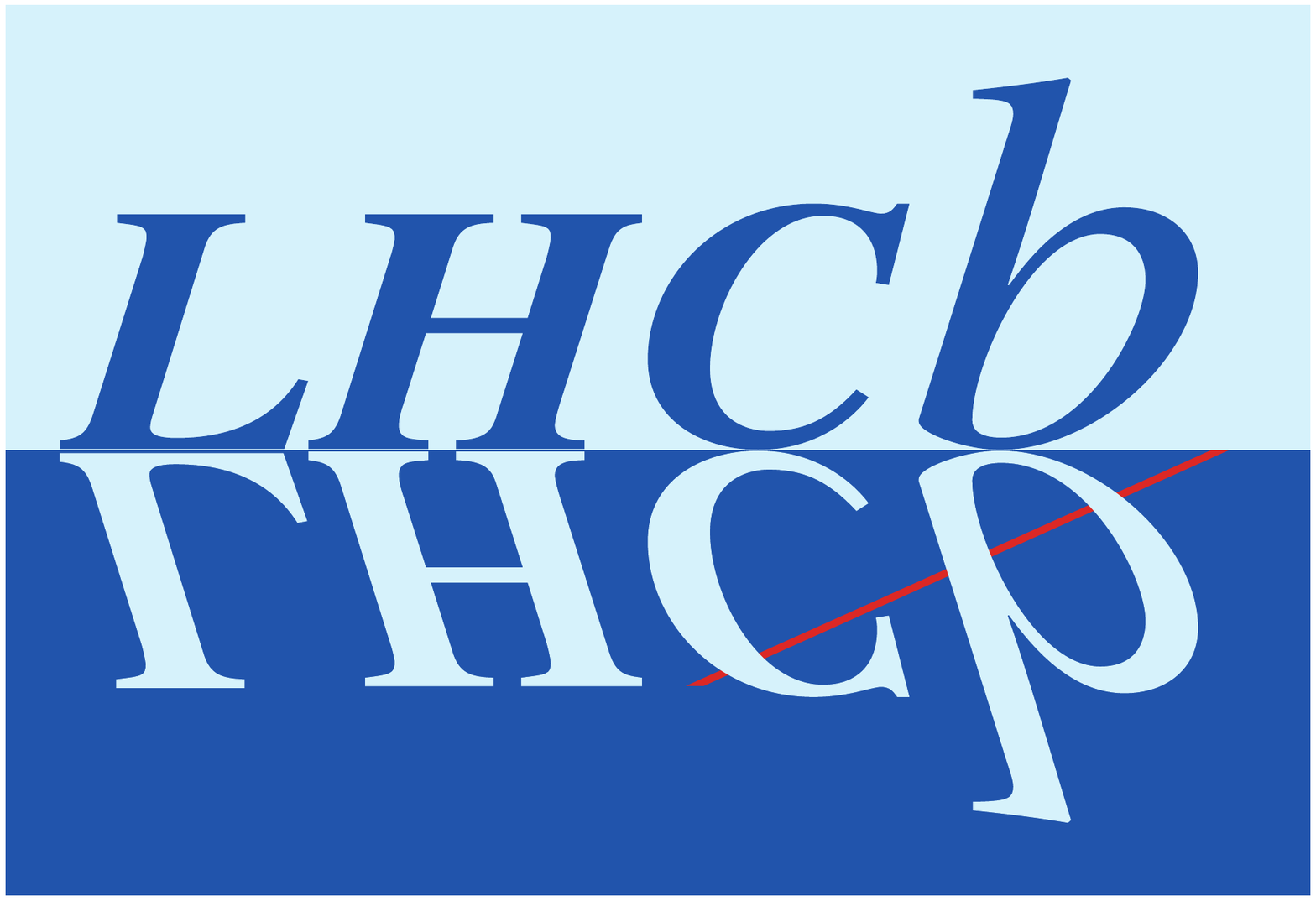}} & &}%
{\vspace*{-1.2cm}\mbox{\!\!\!\includegraphics[width=.12\textwidth]{lhcb-logo.eps}} & &}%
\\
 & & CERN-PH-EP-2015-301 \\  
 & & LHCb-PAPER-2015-049 \\  
 & & \today \\ 
 & & \\
\end{tabular*}

\vspace*{1.0cm}

{\bf\boldmath\huge
\begin{center}
Measurement of forward $W$ and $Z$ boson production in $pp$ collisions \\ at $\sqrt{s} = 8\mathrm{\,Te\kern -0.1em V}$
\end{center}
}

\vspace*{0.15cm}

\begin{center}
The LHCb collaboration\footnote{Authors are listed at the end of this paper.}
\end{center}

\vspace{0.25cm}
\begin{abstract}
\noindent
Measurements are presented of electroweak boson production using data from $pp$ collisions at a centre-of-mass energy of $\sqrt{s} = 8\mathrm{\,Te\kern -0.1em V}$.
The analysis is based on an integrated luminosity of $2.0\mathrm{\,fb}^{-1}$ recorded with the LHCb detector.
The bosons are identified in the $W\rightarrow\mu\nu$ and $Z\rightarrow\mu^{+}\mu^{-}$ decay channels. 
The cross-sections are measured for muons in the pseudorapidity range $2.0 < \eta < 4.5$, with transverse momenta $p_{\rm T} > 20{\mathrm{\,Ge\kern -0.1em V\!/}c}$ and, in the case of the $Z$ boson, a dimuon mass within $60 < M_{\mu^{+}\mu^{-}} < 120{\mathrm{\,Ge\kern -0.1em V\!/}c^{2}}$.
The results are
\begin{align*}
\sigma_{W^{+}\rightarrow\mu^{+}\nu} &= 1093.6 \pm 2.1 \pm 7.2 \pm 10.9 \pm 12.7{\rm \,pb} \, ,\\
\sigma_{W^{-}\rightarrow\mu^{-}\overline{\nu}} &= \phantom{0}818.4 \pm 1.9 \pm 5.0 \pm \phantom{0}7.0 \pm \phantom{0}9.5{\rm \,pb} \, ,\\
\sigma_{Z\rightarrow\mu^{+}\mu^{-}} &= \phantom{00}95.0 \pm 0.3 \pm 0.7 \pm \phantom{0}1.1 \pm \phantom{0}1.1{\rm \,pb} \, ,
\end{align*}
where the first uncertainties are statistical, the second are systematic, the third are due to the knowledge of the LHC beam energy and the fourth are due to the luminosity determination.
The evolution of the $W$ and $Z$ boson cross-sections with centre-of-mass energy is studied using previously reported measurements with $1.0\mathrm{\,fb}^{-1}$ of data at $7\mathrm{\,Te\kern -0.1em V}$.
Differential distributions are also presented.
Results are in good agreement with theoretical predictions at next-to-next-to-leading order in perturbative quantum chromodynamics. 
  
\end{abstract}

\vspace*{0.3cm}

\begin{center}
Published in JHEP 01 (2016) 155
\end{center}

\vspace{\fill}

{\footnotesize 
\centerline{\copyright~CERN on behalf of the \lhcb collaboration, licence \href{http://creativecommons.org/licenses/by/4.0/}{CC-BY-4.0}.}}

\end{titlepage}


\newpage
\setcounter{page}{2}
\mbox{~}

\cleardoublepage


\renewcommand{\thefootnote}{\arabic{footnote}}
\setcounter{footnote}{0}

\pagestyle{plain} 
\setcounter{page}{1}
\pagenumbering{arabic}


\section{Introduction}
\label{sec:Introduction}

Measurements of \W and \Z boson production cross-sections at hadron colliders constitute important tests of the Standard Model (SM).\footnote{Throughout this article \Z is used to denote the $\Z/\gamma^{*}$ contributions.}
Theoretical predictions for these cross-sections are available at next-to-next-to-leading order (NNLO) in perturbative quantum chromodynamics~\cite{nnlopQCD1,nnlopQCD2,nnlopQCD3,nnlopQCD4,nnlopQCD5}. 
The dominant uncertainty on these predictions reflects the uncertainties on the parton density functions (PDFs).
The forward acceptance of the \lhcb detector allows the PDFs to be constrained at Bjorken-$x$ values down to $10^{-4}$~\cite{LHCbPDF}.
Ratios of the \W and \Z cross-sections provide precise tests of the SM as the sensitivity to the PDFs in the theoretical calculations is reduced and many of the experimental uncertainties cancel.

During \lhc \rone, data were collected at centre-of-mass energies, \sqs, of 7\tev and 8\tev, providing two distinct samples for measurements of the electroweak boson production cross-sections.
The evolution of the cross-sections, and cross-section ratios, may be used to infer the existence of physics beyond the Standard Model (BSM)~\cite{COM}.

\lhcb has measured the \W boson production cross-section at \sqs = 7\tev using the muon channel~\cite{wmu}, and that of \Z bosons decaying to muon~\cite{zmu}, electron~\cite{ze} and tau lepton~\cite{ztau} pairs, using a data set of 1.0\invfb. 
The \Z boson production cross-section at \mbox{\sqs = 8\tev} has also been measured using decays to electron pairs~\cite{LHCb-PAPER-2015-003}.
Similar measurements have also been performed by the \atlas~\cite{atlaswz} and \cms~\cite{cmswz,cms8-1,cms8-2} collaborations, although in different kinematic regions.

The measurements of inclusive \W and \Z boson cross-sections at \sqs = 8\tev described here are performed following the same procedure as detailed in Refs.~\cite{wmu,zmu}.
The cross-sections are defined for muons with transverse momentum $\pt > 20\gevc$ and pseudorapidity in the range $2.0 < \eta < 4.5$. 
In the case of the \Z boson measurements, the invariant mass of the two muons is required to be in the range $60 < \mmm < 120\gevcc$. 
These kinematic requirements define the fiducial region of the measurement and are referred to as the fiducial requirements in this article. 
Total cross-sections are presented, as well as differential cross-sections as functions of $\eta$ of the muons, and of the 
\Z boson rapidity, \rapz, transverse momentum, \ptz, and \phist~\cite{phistar}. 
Here \phist is defined as\footnote{The \phist definition in this article is equivalent to the definitions in Refs.~\cite{zmu,ze,LHCb-PAPER-2015-003}.}
\begin{equation}
\phist \equiv \frac{\tan{(\phi_{\rm acop}/2)}}{\cosh{(\Delta\eta}/2)} \, ,
\end{equation}
where the angle $\phi_{\rm acop} = \pi - |\Delta\phi|$ depends on the  
difference $\Delta\phi$ in azimuthal angle between the two muon momenta, 
while the difference between their pseudorapidities is denoted by $\Delta\eta$.
Differential cross-section ratios and the muon charge asymmetry, arising from the W production charge asymmetry, are also determined as a function of the muon pseudorapidity.

This article is organised as follows:
Section~\ref{sec:Detector} describes the \lhcb detector;
Section~\ref{sec:selection} details the selection of \W and \Z boson candidate samples;
Section~\ref{sec:cs} defines the \W and \Z boson cross-sections and summarises the relevant sources of systematic uncertainty, 
as well as their estimation;
Section~\ref{sec:results} presents the results and Section~\ref{sec:conclusions} concludes the article.
Appendices~\ref{sec:differential} and~\ref{sec:correlation} provide tables of differential cross-sections and correlations between these measurements.

\section{Detector and data set}
\label{sec:Detector}

The \lhcb detector~\cite{Alves:2008zz,LHCb-DP-2014-002} is a single-arm forward spectrometer covering the \mbox{pseudorapidity} range $2<\eta <5$, designed for the study of particles containing \bquark or \cquark quarks. 
The detector includes a high-precision tracking system consisting of a silicon-strip vertex detector surrounding the $pp$ interaction region, a large-area silicon-strip detector located upstream of a dipole magnet with a bending power of about $4{\rm\,Tm}$, and three stations of silicon-strip detectors and straw drift tubes placed downstream of the magnet.
The tracking system provides a measurement of momentum, \ptot, of charged particles with a relative uncertainty that varies from 0.5\% at low momentum to 1.0\% at 200\gevc.
The minimum distance of a track to a primary vertex, the impact parameter (\ip), is measured with a resolution of $(15+29/\pt)\mum$, where \pt is the component of the momentum transverse to the beam, in \gevc.
Different types of charged hadrons are distinguished using information from two ring-imaging Cherenkov detectors.
Photons, electrons and hadrons are identified by a calorimeter system consisting of a scintillating-pad detector (SPD), preshower detectors, an electromagnetic calorimeter and a hadronic calorimeter. 
Muons are identified by a system composed of alternating layers of iron and multiwire proportional chambers.
The online event selection is performed by a trigger~\cite{LHCb-DP-2012-004}, which consists of a hardware stage, based on information from the calorimeter and muon systems, followed by a software stage, which applies a full event reconstruction. 
A requirement that prevents events with high occupancy from dominating the processing time of the software trigger is also applied.
This is referred to in this article as the global event cut (GEC).

The measurements presented here are based on $pp$ collision data collected at a centre-of-mass energy of 8\tev, the integrated luminosity amounting to $1978 \pm 23\invpb$.
The absolute luminosity scale was measured during dedicated data-taking periods, using both van der Meer scans~\cite{vandermeer}  and beam-gas imaging methods~\cite{beamgas}. 
Both methods give consistent results, which are combined to give the final luminosity estimate with an uncertainty of 1.16\%~\cite{lumi2}. 

Several samples of simulated events are produced to estimate contributions from background processes, to verify efficiencies and to correct data for detector-related effects. 
In the simulation, \pp collisions are generated using \pythia8~\cite{Sjostrand:2007gs,Sjostrand:2006za} with 
a specific \lhcb configuration~\cite{LHCb-PROC-2010-056}.
Decays of hadronic particles are described by \evtgen~\cite{Lange:2001uf}, in which final-state radiation is generated using \photos~\cite{Golonka:2005pn}.
The interaction of the generated particles with the detector, and its response, are implemented using the \geant toolkit~\cite{Allison:2006ve,Agostinelli:2002hh} as described in Ref.~\cite{LHCb-PROC-2011-006}.

The \W boson yields are determined from fits to the data using signal templates produced with the 
\resbos~\cite{GEN-RESBOS1,GEN-RESBOS2,GEN-RESBOS3} generator configured with the CT14~\cite{PDF-CT14} PDF set. 
The \resbos generator includes an approximate NNLO calculation, plus a next-to-next-to-leading logarithm approximation for the resummation of the soft gluon radiation.
 
The results of the analysis are compared to theoretical predictions calculated with the \fewz~\cite{fewz,fewzold} generator at NNLO for the PDF sets ABM12~\cite{abm12}, CT10~\cite{ct10}, CT14, HERA1.5~\cite{h1zeus}, MMHT14~\cite{mmht14} and NNPDF3.0~\cite{PDF-NNPDF30}. 
All calculations are performed with the renormalisation and factorisation scales set to the electroweak boson mass.
Scale uncertainties are estimated by varying these scales by factors of two around the boson mass~\cite{scaleUnc}.
Total uncertainties correspond to those coming from the PDF and the strong force coupling strength, \as, both at 68.3\% confidence level (CL), added in quadrature with the scale 
uncertainties.

\section{Event yield}
\label{sec:selection}

Events for this analysis must satisfy the selection criteria detailed in Refs.~\cite{wmu,zmu}.
The trigger requires a single muon with $\pt > 1.5\gevc$ at the hardware stage, and includes an upper threshold of 600 hits in the SPD to prevent high-particle multiplicity events from dominating the processing time. 
A muon with $\pt > 10\gevc$ is required at the software stage. 
In the offline analysis, particles are required to be well-reconstructed, to be identified as muons, and also to pass the fiducial requirements. 

Additional selection criteria are applied to the \W boson candidates to reduce the contributions of various sources of background. 
Muons from decays of \W bosons are generally isolated from other particles. 
To define a degree of isolation, a cone with radius \mbox{$R=\sqrt{\Delta\eta^{2} + \Delta\phi^{2}} = 0.5$} is constructed around the direction of the muon track.
Excluding the candidate muon momentum, requiring that there are small amounts of transverse momentum ($\ptcone < 2\gevc$) and transverse energy ($\etcone < 2\gev$) in the cone reduces background originating from generic QCD events. 
Requiring the transverse momentum of all other muons in the event to be less than 2\gevc reduces the contamination from \zmm events.  
An upper limit on the \ip of 40\mum removes candidates in which the muon is not consistent with originating from the primary vertex. 
Such candidates could be due to electroweak boson decays to tau leptons, which in turn decay to muons, or semileptonic decays of heavy flavour hadrons.
Genuine muons are expected to leave low-energy deposits in the electromagnetic and hadronic calorimeters. 
An upper limit of 4\% on the amount of energy that is deposited in the calorimeters relative to the momentum of the track (\eop) reduces the background from energetic pions and kaons punching through the calorimeters to the muon stations.
A total of $1\,733\,327$ \wmn candidates are identified.

\begin{figure}[t]
\begin{center}
\includegraphics[width=0.49\textwidth]{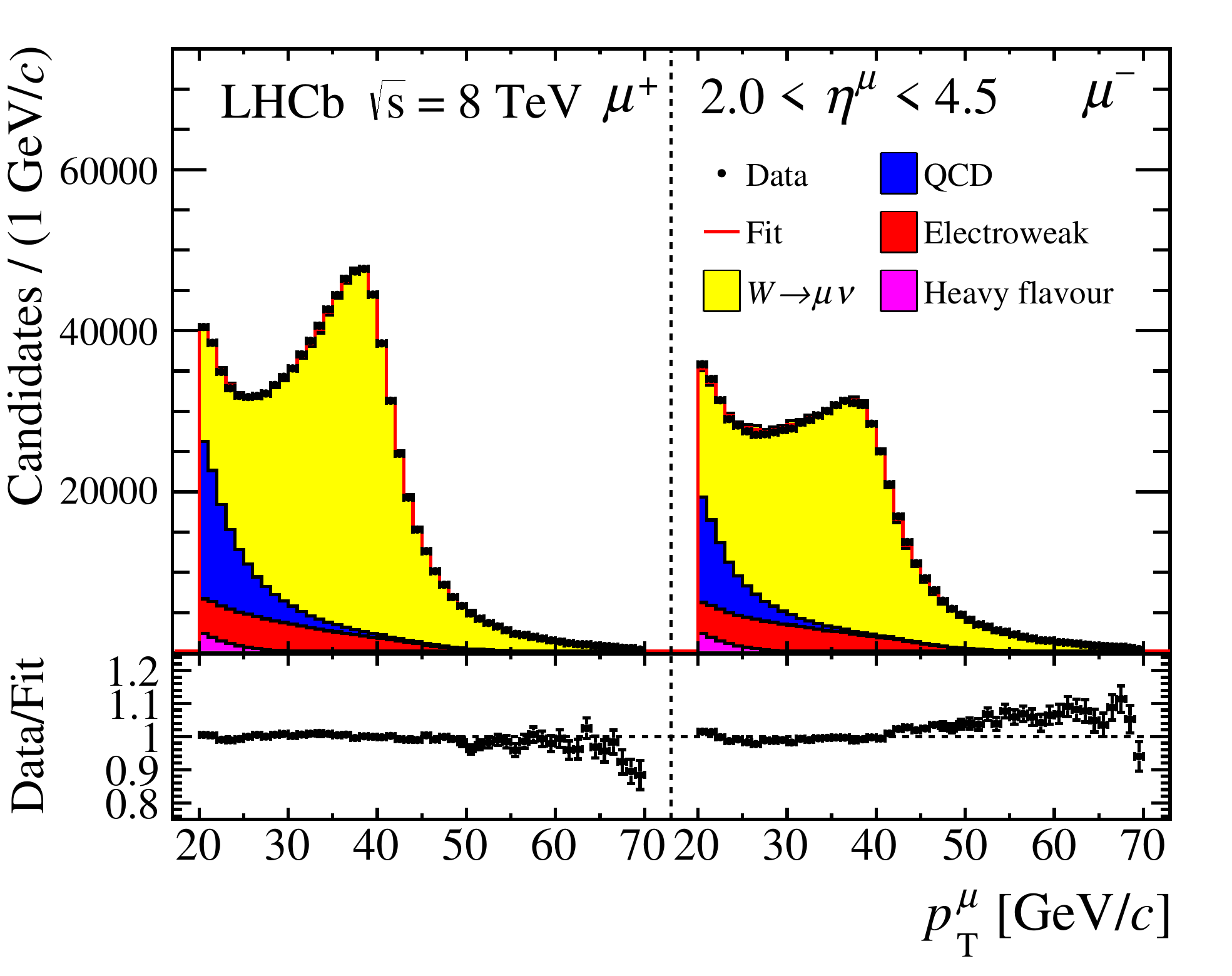}
\includegraphics[width=0.49\textwidth]{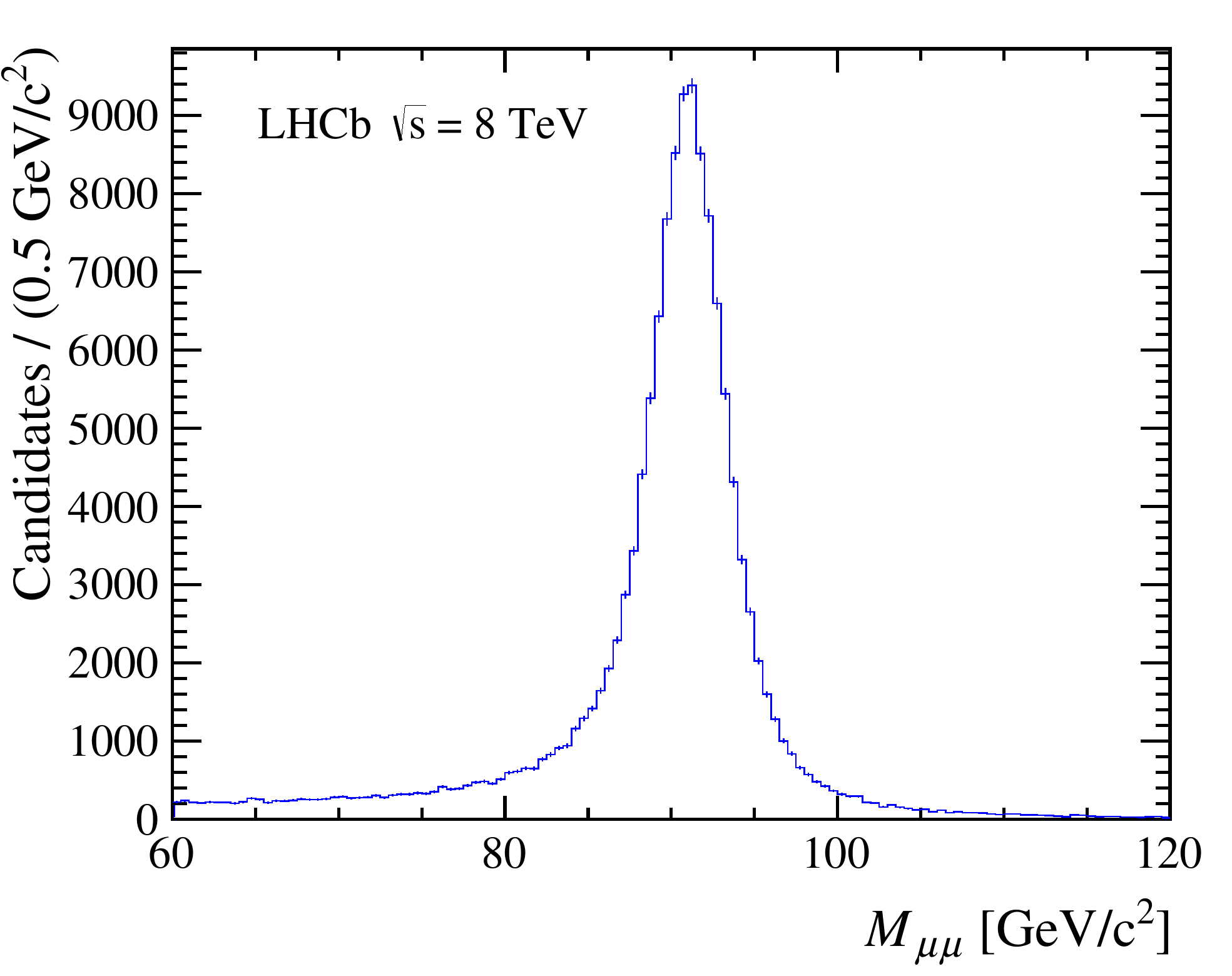}
\caption{(left) Template fit to the (left panel) positive and (right panel) negative muon \pt spectra in the full $\etamu$ range for \W candidates. Data are compared to fitted contributions from \wmn signal and QCD, electroweak and heavy flavour backgrounds. (right) Invariant mass distribution of dimuon pairs in the \Z candidate sample.} 
\label{fig:yields}
\end{center}
\end{figure}

The \Wpm sample purity ($\rho^{\Wpm}$), defined as the ratio of signal to candidate event yield, is determined with a template 
fit to 
the positively and negatively charged muon \pt distributions in eight bins of muon pseudorapidity using the method of extended maximum likelihood.	
Only muons with \pt smaller than 70\gevc are considered. 
The \Wpm boson signal and the \zmm background templates are based on distributions predicted by the \resbos generator. 
The \ztt and \wtn templates are taken from \pythia8 simulation. 
The overall fraction of the electroweak background (\zmm, \ztt and \wtn decays) in the \W boson candidate sample is determined using a data-driven method to be $(10.84 \pm 0.21)\%$. 
A template for backgrounds due to misidentified hadrons is taken from data using a sample of randomly triggered pions and kaons that are weighted by their probability to be misidentified as muons.
This component is left free to vary in the fit, and is determined to account for about 9.6\% of the total candidates.
Finally, a template of heavy-flavour decays is obtained from data using muons with an \ip of more than 100\mum. 
The fraction of this background is determined from a fit to the muon \ip distribution and is found to be $(1.31 \pm 0.09)\%$ of 
the \W boson candidate sample.
The momentum calibration for high-\pt muons is performed using the data-driven technique outlined in Ref.~\cite{LHCb-PAPER-2015-039}.
A more detailed description of the fit implementation is given in Ref.~\cite{wmu}.
The fit result in the full $\etamu$ range is presented in Fig.~\ref{fig:yields} (left), where the normalised residuals show an imperfect description of the data by the adopted templates, 
similar to the 7\tev analysis.
The effect of this discrepancy on the signal yield is at the few per mille level.
The overall purities are $\rho^{\Wp} = (78.91 \pm 0.15)\%$ and $\rho^{\Wm} = (77.49 \pm 0.18)\%$.

The invariant mass distribution of dimuon pairs passing the \Z candidate requirements is shown in Fig.~\ref{fig:yields} (right).
In total, $136\,702$ \zmm candidates are selected.
The background contamination is low.
Five background sources are considered: decays of heavy flavour hadrons, hadron misidentification, \ztt decays, \ttbar and \W\W production.
The largest sources of background are due to decays of heavy flavour hadrons
and hadron misidentification. 
These backgrounds are determined from data using the techniques discussed in Ref.~\cite{zmu}. 
The heavy-flavour background is estimated from a subset of the candidate sample by placing additional requirements on muon isolation and dimuon vertex quality. 
The background due to hadron misidentification is estimated using pairs of hadrons from randomly triggered data. 
These are weighted by the momentum-dependent probabilities for hadrons to be misidentified as muons.
The other backgrounds are determined using simulation and the purity is measured to be $\rho^{Z} = (99.3 \pm 0.2)\%$. 

\section{Cross-section measurement}
\label{sec:cs}

Cross-sections are determined in the specified kinematic ranges and are corrected for quantum electrodynamic (QED) final-state radiation (FSR) in order to compare measurements of electroweak boson production in different decay modes and to provide a consistent comparison with next-to-leading order and NNLO QCD predictions.
No corrections are applied for initial-state radiation or for electroweak effects and their interplay with QED effects.

The \W boson cross-sections are measured as a function of \etamu using the equation
\begin{equation}
\label{eq:csw}
\sigma_{\Wpm \rightarrow \mu^{\pm} \nu }(i) = 
\frac{\rho^{\Wpm}(i)}{\lum} \cdot
\frac{\ffsr^{\Wpm}(i)}{\effgec(i)} \cdot
\frac{N^{\Wpm}(i)}{\varepsilon^{\Wpm}(i) \, \effsel^{\Wpm}(i) \, \mathcal{A}^{\Wpm}(i)} \, ,
\end{equation}
\noindent where all quantities except for the integrated luminosity, \lum, are determined in each bin $i$ of \etamu.
The number of observed \Wpm boson candidates is denoted by $N^{\Wpm}(i)$.
The correction factors for QED final-state radiation are given by $\ffsr^{\Wpm}$ and the efficiency of the requirement on the number of SPD hits in the hardware trigger is represented by \effgec.
The total muon reconstruction efficiency is denoted by $\varepsilon^{\Wpm}$ while the efficiency of the selection criteria is given by $\effsel^{\Wpm}$.
The acceptance correction, $\mathcal{A}^{\Wpm}$, accounts for the 70\gevc experimental upper bound in the fit to the muon \pt. 

The \Z boson cross-sections are measured in bins of \rapz, \ptz, \phist and \etamu, by integrating over all but one of these variables.
To account for bin migration effects, the cross-section in bin $i$ is determined from the number of events in all bins $j$ with an unfolding matrix $U$, as follows
\begin{equation}
\label{eq:csz}
\csz(i) = 
\frac{\rho^{Z}}{\lum} \cdot
\frac{\ffsr^{\Z}(i)}{\effgec(i)} \cdot
\sum_j U(i,j) \left( \sum_k \frac{1}{\varepsilon^{Z}(\eta_{k}^{\mu^{+}},\eta_{k}^{\mu^{-}})}\right)_{j} \, .
\end{equation}
In this expression the index $k$ runs over all candidates contributing to bin $j$ and $\varepsilon^{\Z}$ is the pseudorapidity-dependent muon-reconstruction efficiency for event $k$.
The matrix $U$ is determined from simulated data, as described in Section~\ref{sec:unfld}.
The QED final-state radiation corrections are denoted by $\ffsr^{Z}$.
The components that are common with the \W boson cross-sections defined in Eq.~\ref{eq:csw} are the luminosity and the individual muon reconstruction efficiencies.
 
Although the beam energy does not enter in Eqs.~\ref{eq:csw} and \ref{eq:csz}, a related uncertainty is assigned to all cross-sections.
More details on these individual components are given below.

\subsection{Muon reconstruction efficiencies}

The data are corrected for inefficiencies associated with track reconstruction, muon identification, and trigger requirements. 
All efficiencies are determined from data using the techniques detailed in Refs.~\cite{wmu,zmu}, where the track reconstruction, muon identification, and muon trigger efficiencies are obtained using tag-and-probe methods applied to the \Z candidates. 
The tag and the probe tracks are required to satisfy the fiducial requirements.
The tag must be identified as a muon and be consistent with triggering the event, while the probe is defined so that it is unbiased with respect to the requirement for which the efficiency is being measured.
The efficiency is studied as a function of several variables, which include both the muon momenta and the detector occupancy. In this analysis, reconstruction, identification and trigger efficiencies are applied as a function of the muon pseudorapidity. 
The efficiency in each bin of \etamu is defined as the fraction of tag-and-probe candidates where the probe satisfies the corresponding track reconstruction, identification or trigger requirement. 
All efficiencies are observed to be independent of the muon charge.

The tracking efficiency is determined using probe tracks that are reconstructed by combining hits from the muon stations and the large-area silicon-strip detector. 
The muon identification efficiency is determined using probe tracks that are reconstructed without using the muon system.
The single-muon trigger efficiency is determined using reconstructed muons 
as probes.

\subsection{GEC efficiency}
\label{sec:gec}

The efficiency of the SPD multiplicity limit at 600 hits in the muon trigger is evaluated from data using two independent methods. 
The first exploits the fact that the SPD multiplicities of single $pp$ interactions involving a \Z boson are rarely 
above the 600 hit threshold.
The expected SPD multiplicity distribution of signal events is constructed by adding the multiplicities of signal events in single $pp$ interactions to the multiplicities of randomly triggered events, as in Ref.~\cite{lhcbwz}.
The convolution of the distributions extends to values above 600 hits, and the fraction of events that the trigger rejects can be determined.
The second method consists of fitting the SPD multiplicity distribution and extrapolating the fit function to determine the fraction of events that are rejected,
as in Ref.~\cite{zmu}. 
Both methods give consistent results and $\effgec = (93.00 \pm 0.32)\%$ is used in this analysis as the overall efficiency. 
This efficiency depends linearly on \rapz and \etamu, with about 2\% variation across the full range. 
This is accounted for by applying a bin-dependent efficiency correction.

\subsection{Final-state radiation}

The FSR correction is taken to be the mean of the corrections calculated with \herwigpp~\cite{herwig} and \pythia8.
The corrections are tabulated in Appendix~\ref{sec:differential} and are about 2.5\% on average.

\subsection{Selection efficiencies}

The efficiency of the additional selection requirements for the \W boson candidate samples is evaluated using a sample of \Z 
bosons from data, where one of the muons is excluded to mimic a \wmn decay~\cite{wmu}.
However, this introduces a bias because the \pt distribution of muons from \Z bosons is harder than those from \W bosons.
Simulation is used to correct for this bias and for the fact that the \Z boson sample requires two muons in the LHCb acceptance.

\subsection{Acceptance}

Only muons with \pt smaller than 70\gevc are considered for the extraction of the \W boson signal. 
A kinematic acceptance correction is required in order to measure cross-sections without this restriction on muon \pt. 
This correction is evaluated using the \resbos simulated sample.

\subsection{Unfolding the detector response}
\label{sec:unfld}

To correct for detector resolution effects, an unfolding is performed (matrix $U$ of Eq.~\ref{eq:csz}) using \lhcb simulation and the \roounfold~\cite{roounfold} software package.
Only the \ptz and \phist distributions are unfolded. 
Since \rapz and \etamu are well measured, no unfolding is performed. 
The momentum resolution in the simulation is calibrated to the data.
The data are then unfolded using the iterative Bayesian approach proposed in Ref.~\cite{dagostini}.
Other unfolding techniques~\cite{svd,cowan} give similar results.
Additionally, all unfolding methods are tested for model dependence using underlying distributions from leading-order \pythia8, leading-order \herwigpp, as well as next-to-leading order \powheg~\cite{powheg, powheg1, powheg2} showered with both \pythia8 and \herwigpp using the \powheg matching scheme. 
The corrections are between 0.5--8.0$\%$ as a function of \ptz, and between 0.1--7.0$\%$ as a function of \phist.

\subsection{Systematic uncertainties}
\label{sec:syst}

Sources of systematic uncertainty and their effects on the total cross-section measurements from the \sqs = 8\tev data set are summarised in Table~\ref{tab:syst}.
The uncertainty due to the momentum correction is negligible.
Uncertainties from external input, \eg the beam energy and luminosity determinations, are quoted separately from the other contributions.

\begin{table}[!t]
\begin{center}
\caption{Summary of the relative uncertainties on the \Wp, \Wm and \Z boson cross-sections.}
\label{tab:syst}
\begin{tabular}{cccc}
Source & \multicolumn{3}{c}{Uncertainty [\%]} \\ \hline
& \cswp & \cswm & \csz  \\ \hline
Statistical & $0.19$ & $0.23$ & $0.27$ \\ \hline
Purity & $0.28$ & $0.21$ & $0.21$ \\ 
Tracking & $0.26$ & $0.24$ & $0.48$ \\
Identification & $0.11$ & $0.11$ & $0.21$ \\
Trigger & $0.14$ & $0.13$ &  $0.05$\\
GEC & $0.40$ & $0.41$ & $0.34$ \\
Selection & $0.24$ & $0.24$ & --- \\
Acceptance and FSR & $0.16$ & $0.14$ & $0.13$ \\ \hline
Systematic & $0.65$ & $0.61$ &  $0.67$ \\ \hline
Beam energy & $1.00$ & $0.86$ & $1.15$ \\
Luminosity & $1.16$ & $1.16$ & $1.16$ \\ \hline
Total & $1.67$ & $1.59$ & $1.79$ \\
\end{tabular}
\end{center}
\end{table}

For the \W boson samples, the systematic uncertainty on the purity is estimated by considering different shapes and normalisations of the templates, refitting, and summing in quadrature the largest observed deviation in the results corresponding to each source~\cite{wmu}. 
The uncertainty on the \resbos signal template shape includes the effects of the PDF, the factorisation scale and 
the renormalisation scale.
An alternative definition for the QCD background template, potential mismodelling of the lepton \pt shape in \pythia for events that contain jets, and the normalisations of the background templates are accounted for with additional uncertainties.
The total uncertainties on the \Wp and \Wm integrated cross-sections from the sample purity are 0.28\% and 0.21\%.
For the \Z boson sample, the systematic uncertainty on the purity is determined by considering alternative definitions of the heavy-flavour background samples, and by varying by their uncertainties the probabilities for hadrons to be misidentified as muons.
In addition, an uncertainty accounting for the assumption that the purity is the same for all variables and bins of the analysis is evaluated by comparing to cross-section measurements using a binned purity, rather than a global one. 
The uncertainties on the differential cross-section measurements due to variations in purity are typically less than 1\%.

The systematic uncertainty associated with the trigger, identification and tracking efficiencies is determined by re-evaluating all cross-sections with the values of the individual efficiencies increased or decreased by one standard deviation. 
The full covariance matrix of the differential cross-section measurements is evaluated in this way for each source of uncertainty. 
The covariance matrices for each source are added and the diagonal elements of the result determine the total systematic uncertainty due to reconstruction efficiencies.
The total uncertainties on the \Wp, \Wm and \Z boson integrated cross-sections due to reconstruction efficiencies are 0.32\%, 
0.29\% and 0.53\%.

The GEC efficiency for events containing a \Z boson is $\effgec^{Z} = (93.00 \pm 0.32)\%$.
Differences between this efficiency and those for events containing a \Wp or \Wm boson are expected to be small, 
and are thus accounted for with additional systematic uncertainties, as explained in Ref.~\cite{zmu}.
The values used for the measurements of \W boson cross-sections are $\effgec^{W^{+}} = (93.00 \pm 0.37)\%$ and \mbox{$\effgec^{W^{-}} = (93.00 \pm 0.38)\%$}.

The uncertainties due to \W boson selection efficiencies result in uncertainties on the \Wp and \Wm integrated cross-sections 
of 0.24\% and 0.23\%.
These include the uncertainties that arise due to the difference in \W and \Z boson muon \pt spectra and the correction that accounts for the fact that two muons are required to be inside the LHCb acceptance in the \Z boson data sample.

As an estimate of the uncertainty due to the acceptance correction, half the difference between the corrections evaluated using the \resbos generators and \pythia8 is taken. 
This results in uncertainties on the \Wp and \Wm integrated cross-sections of 0.06\% and 0.09\%.

The systematic uncertainty on the FSR correction is the quadratic sum of two components. The first is due to the statistical precision of the 
\pythia8 and \herwigpp estimates and the second is 
half of the difference between their central values, where the latter dominates.

The measurements are specified at a \pp centre-of-mass energy of \sqs = 8\tev.
The beam energy, and consequently the centre-of-mass energy, is known to 0.65\%~\cite{BEAM}. 
The sensitivity of the cross-section to the centre-of-mass energy is studied with the 
\dynnlo~\cite{dynnlo} generator at \nnlo.
Cross-sections are calculated at 1~TeV intervals in centre-of-mass energy and a functional form for the cross-section is determined from a spline interpolation.
A 0.65\% uncertainty on the centre-of-mass energy induces relative uncertainties of 1.00\%, 0.86\% and 1.15\% on the expected \Wp, \Wm and \Z cross-sections.

The uncertainty on the luminosity determination is 1.16\%~\cite{lumi2}, which represents the largest contribution to the total uncertainty.

\section{Results} 
\label{sec:results}

\subsection{Cross-sections at {\boldmath \sqs} = 8 TeV}
\label{sec:wzresults}

\begin{figure}[!t]
\begin{center}
\includegraphics[width=0.84\textwidth]{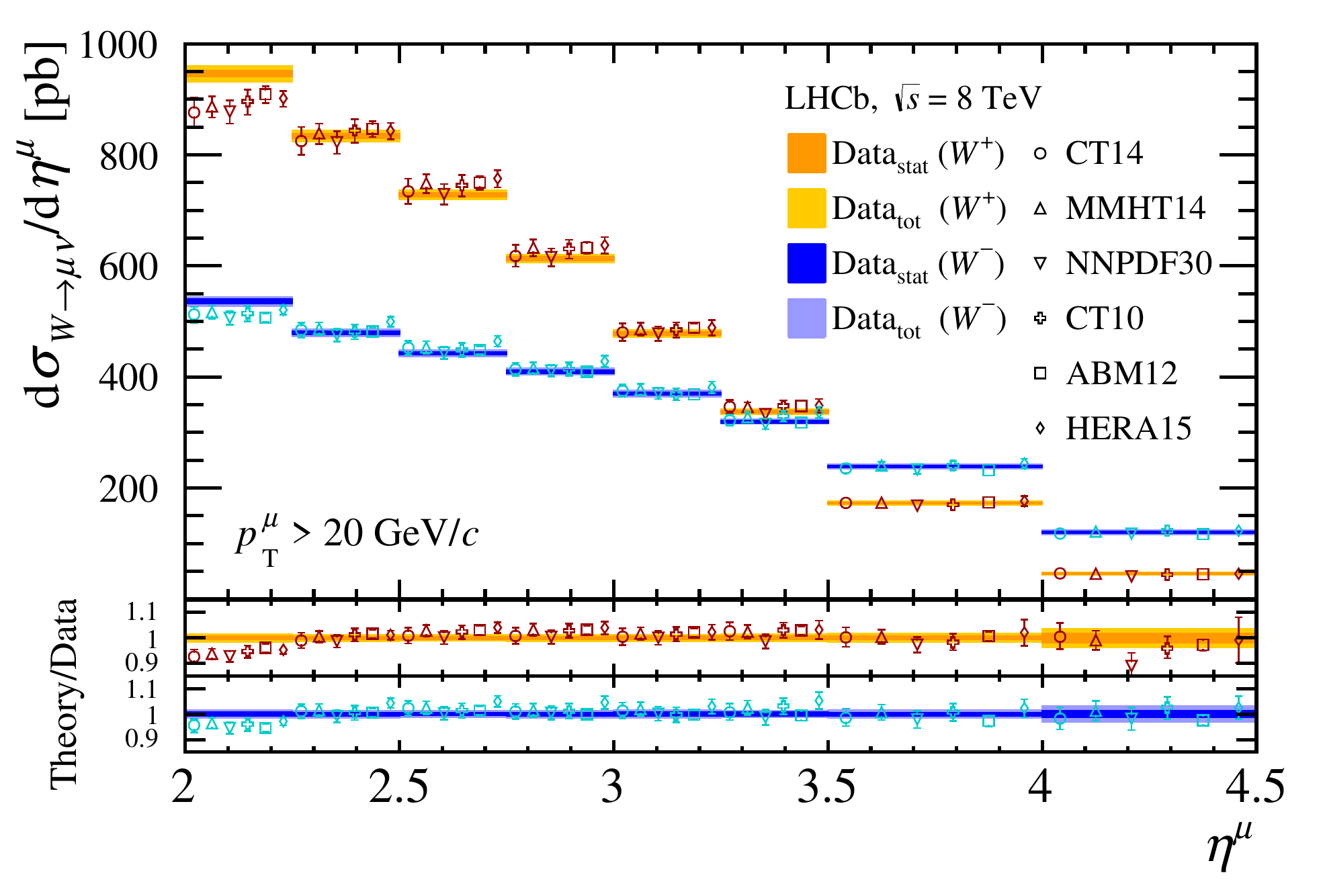}
\includegraphics[width=0.84\textwidth]{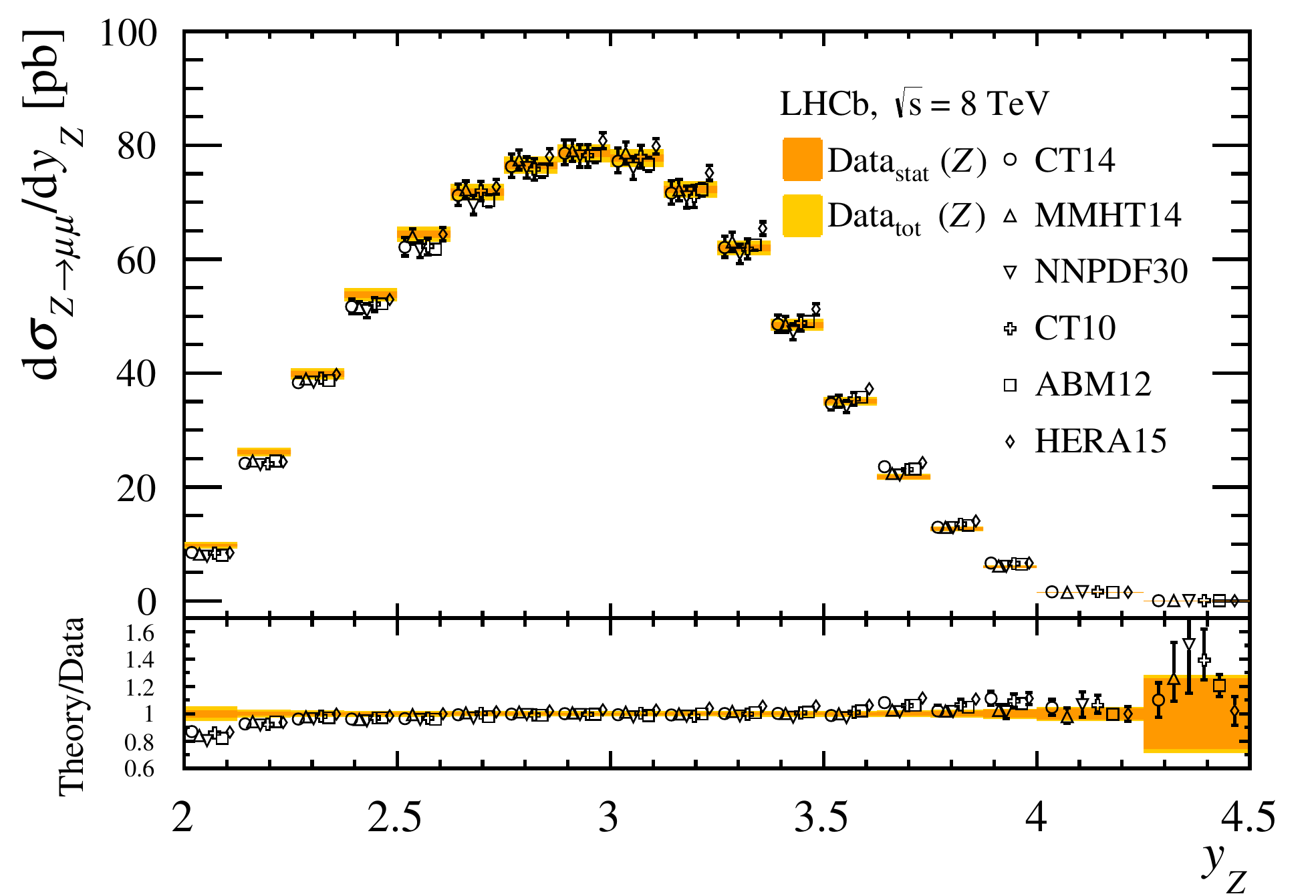}
\caption{(top) Differential \Wp and \Wm boson production cross-section in bins of muon pseudorapidity. (bottom) Differential \Z boson production cross-section in bins of boson rapidity. Measurements, represented as bands, are compared to (markers, displaced horizontally for presentation) NNLO predictions with different parameterisations of the PDFs.}
\label{fig:cs}
\end{center}
\end{figure}

The measured cross-section as a function of muon pseudorapidity in \W boson decays is shown in Fig.~\ref{fig:cs} (top).
Good agreement with the predictions of the \fewz generator, with six different PDF sets, is observed. 
Similar conclusions can be drawn from the comparisons of \Z boson cross-section measurements with predictions as a function of rapidity, as shown in Fig.~\ref{fig:cs} (bottom).
All differential cross-sections are detailed in Tables~\ref{tab:csWETA},~\ref{tab:csZY},~\ref{tab:csZPT},~\ref{tab:csZPHI} and~\ref{tab:csZETA} of Appendix~\ref{sec:differential}.

\begin{figure}[!t]
\begin{center}
\includegraphics[width=0.8\textwidth]{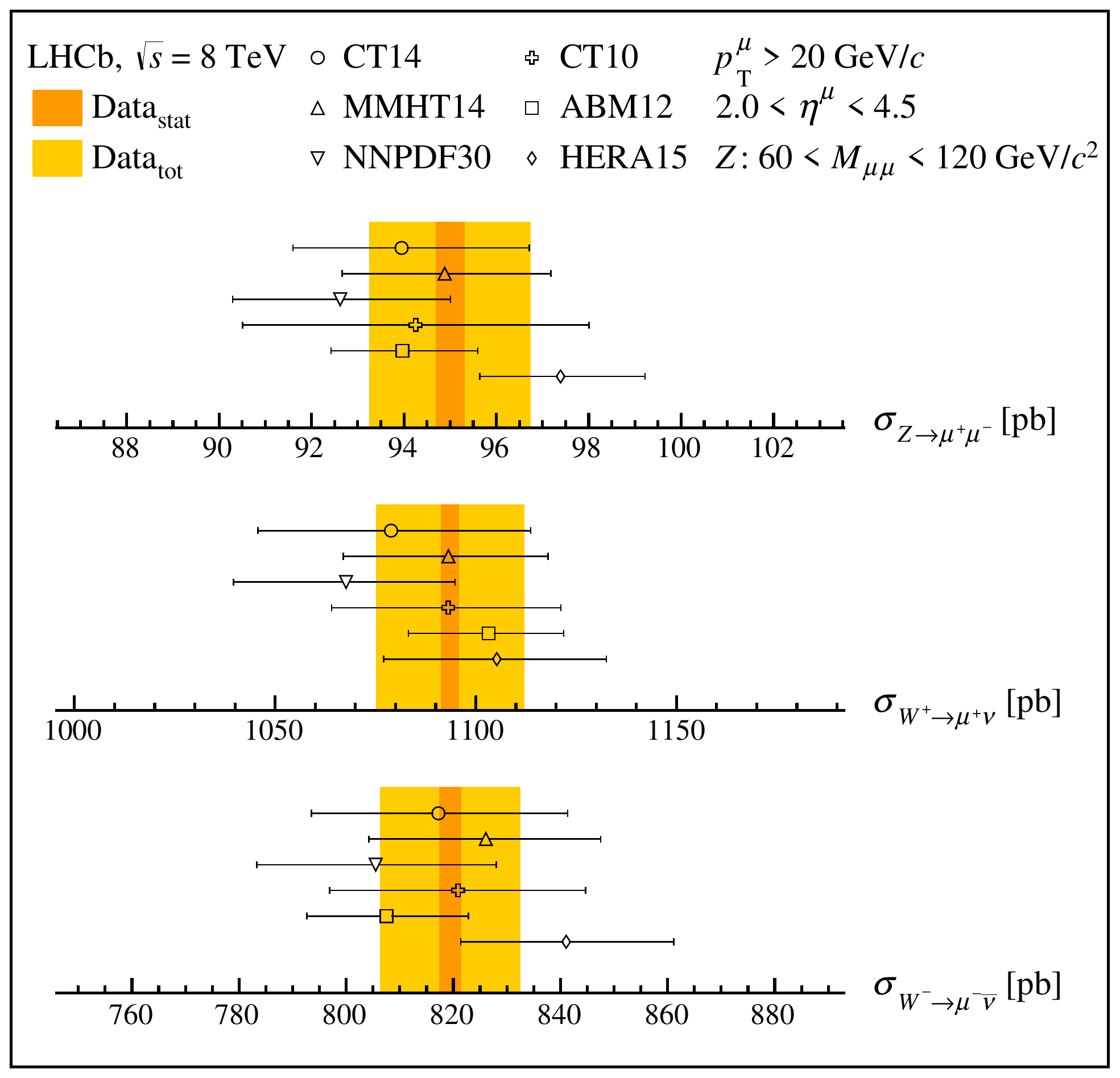}
\caption{Summary of the \W and \Z cross-sections. Measurements, represented as bands, are compared to (markers) \nnlo predictions with different parameterisations of the \pdfs.}
\label{fig:summary_cs}
\end{center}
\end{figure}

\begin{figure}[!t]
\begin{center}
\includegraphics[width=0.49\textwidth]{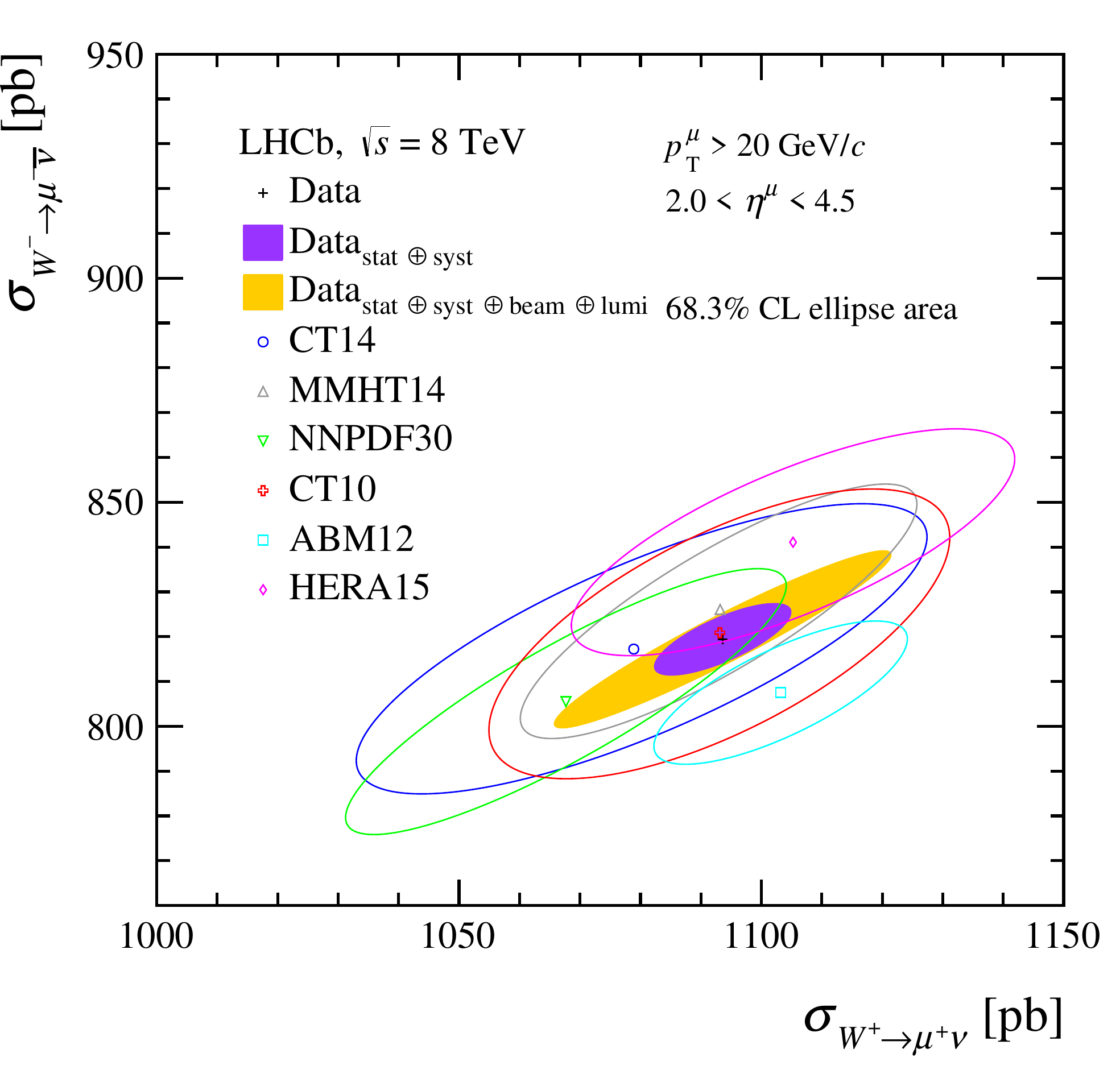}
\includegraphics[width=0.49\textwidth]{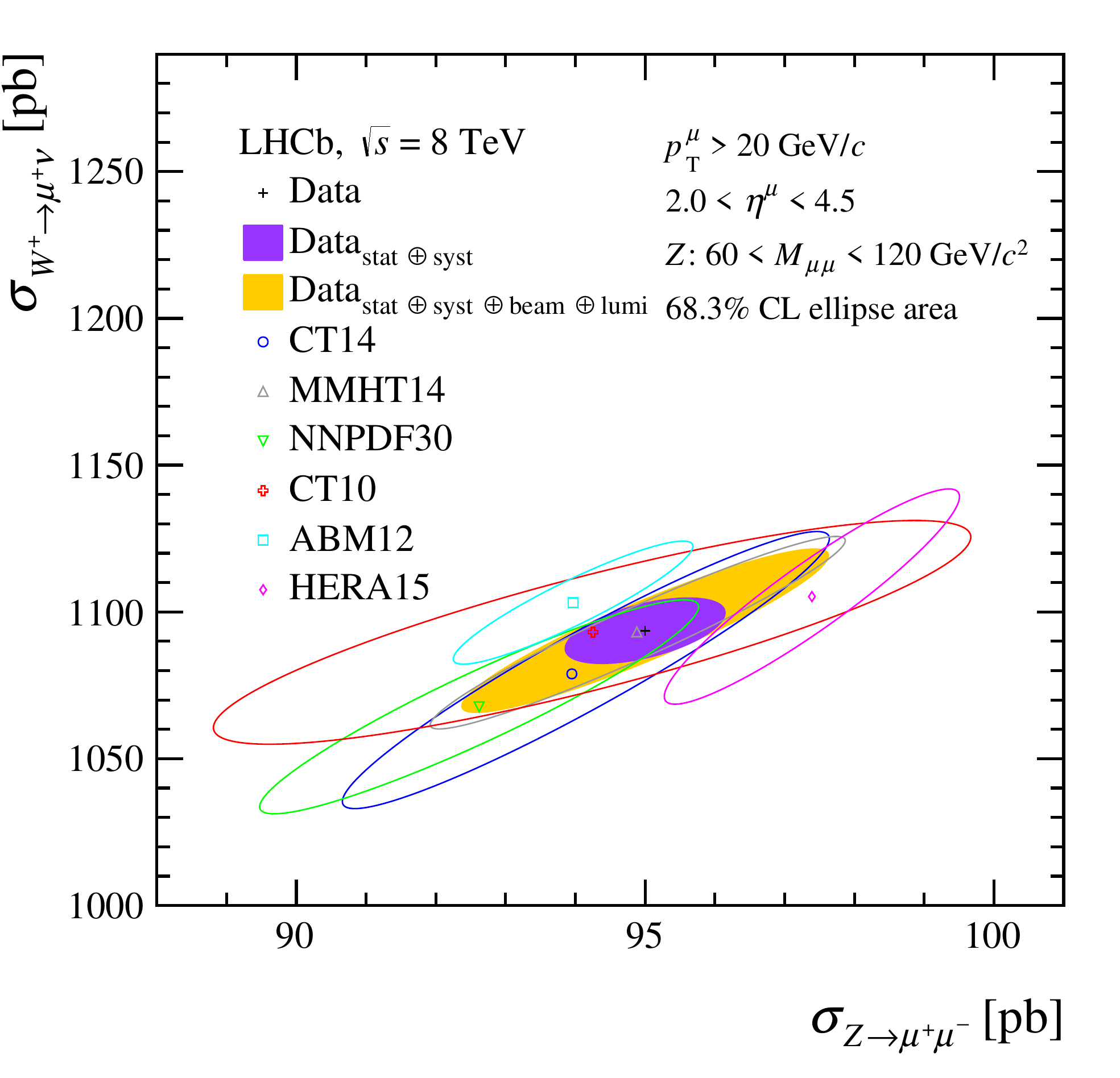} \\
\includegraphics[width=0.49\textwidth]{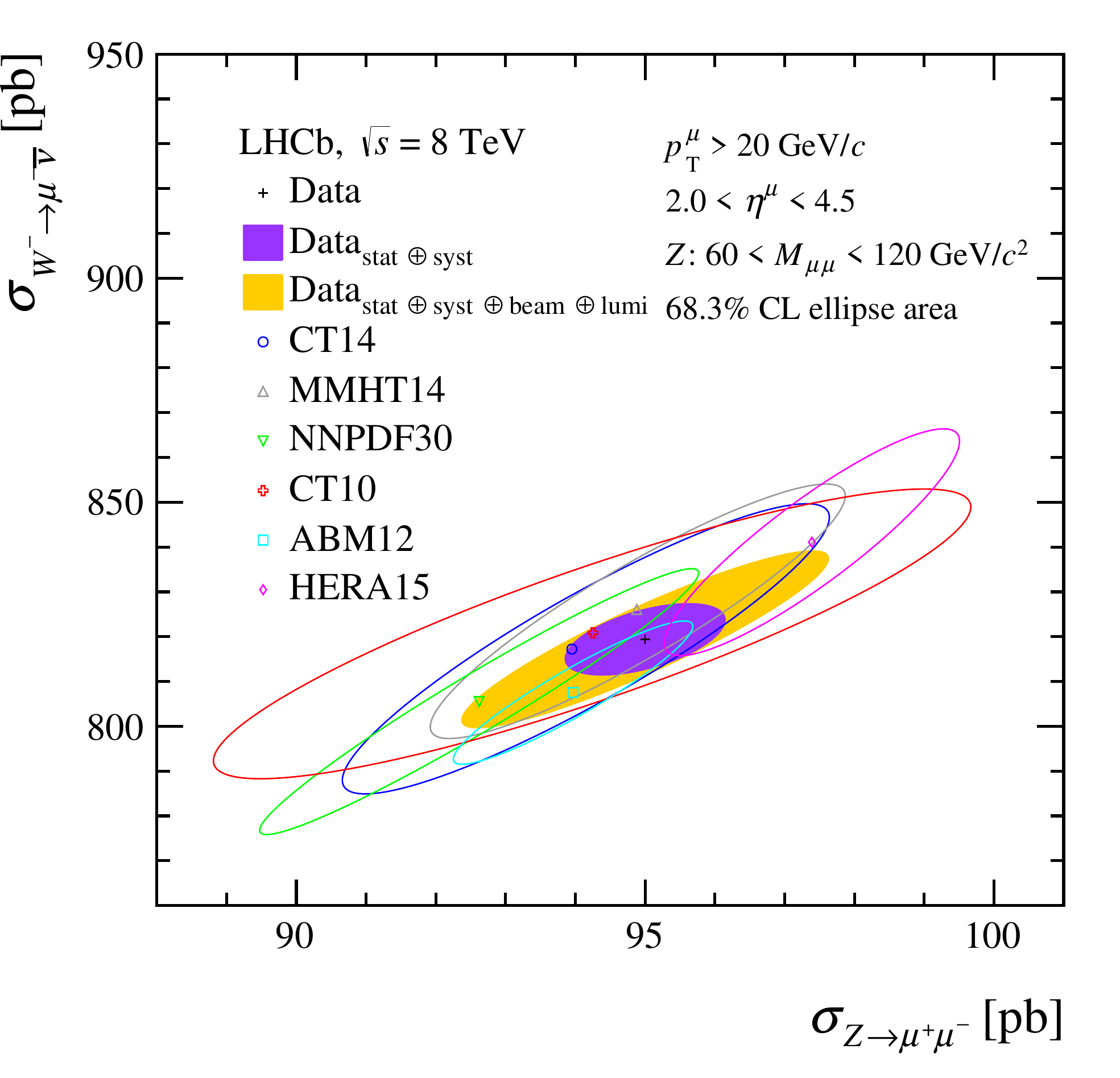}
\includegraphics[width=0.49\textwidth]{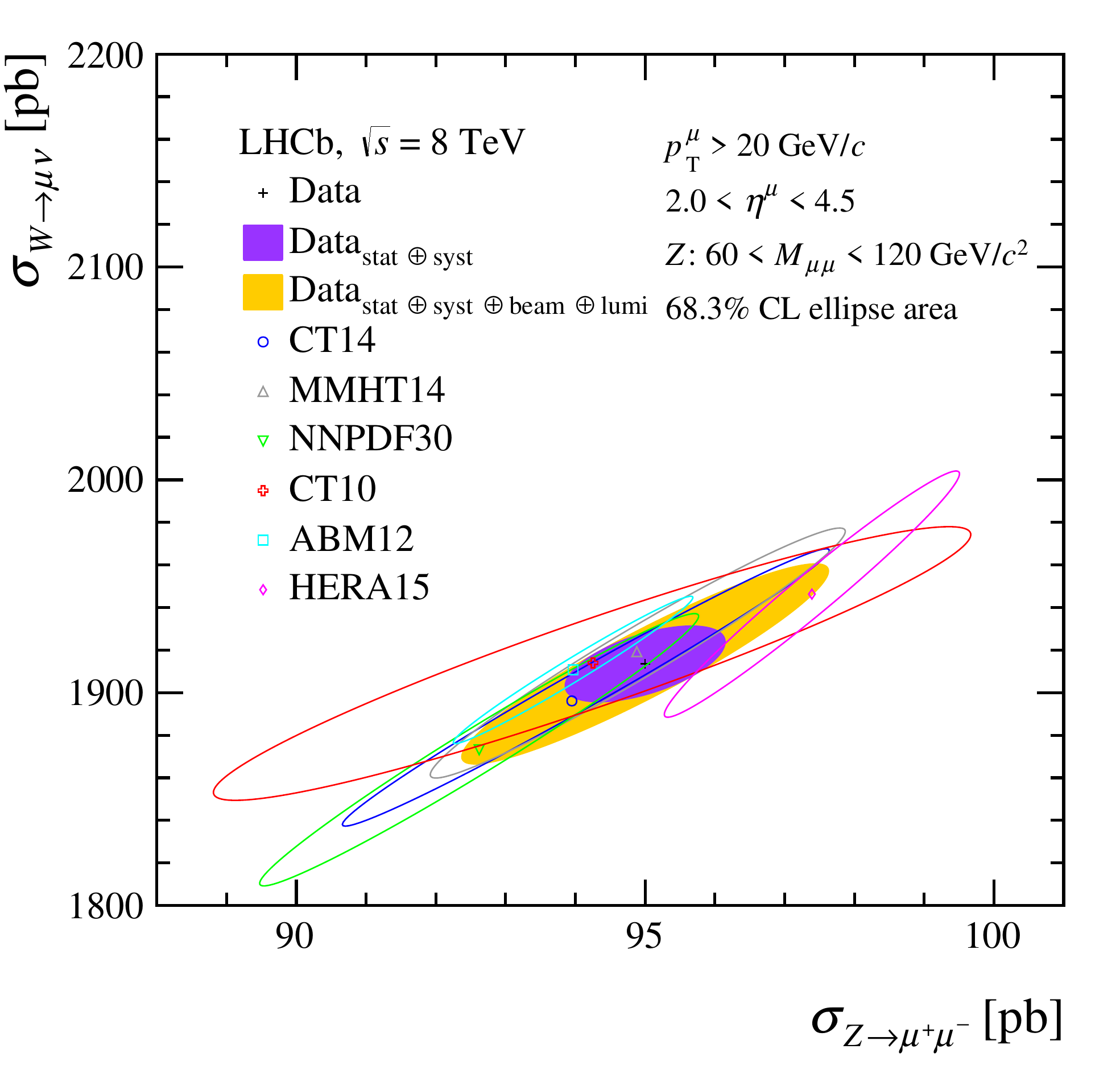}
\caption{Two-dimensional plots of electroweak boson cross-sections compared to NNLO predictions for various parameterisations of the PDFs. The uncertainties on the theoretical predictions correspond to the PDF uncertainty only. All ellipses correspond to uncertainties at 68.3$\%$ CL.}
\label{fig:ellipses}
\end{center}
\end{figure}

The total cross-sections are measured to be
\begin{align*}
\cswp &= 1093.6 \pm 2.1 \pm 7.2 \pm 10.9 \pm 12.7\pb \, ,\\
\cswm &= \phantom{0}818.4 \pm 1.9 \pm 5.0 \pm \phantom{0}7.0 \pm \phantom{0}9.5\pb \, ,\\
\csz &= \phantom{00}95.0 \pm 0.3 \pm 0.7 \pm \phantom{0}1.1 \pm \phantom{0}1.1\pb \, ,
\end{align*}
\noindent where the first uncertainties are statistical, the second are systematic, the third are due to the knowledge of the \lhc beam energy and the fourth are due to the luminosity measurement.
The agreement of the measurements with NNLO predictions given by the \fewz generator configured with various PDF sets is illustrated in Fig.~\ref{fig:summary_cs}.
Two-dimensional plots of electroweak boson cross-sections are shown in Fig.~\ref{fig:ellipses}, where the ellipses correspond to 68.3\% CL coverage.

A best linear unbiased estimator~\cite{BLUE} is used to combine the \Z boson production cross-section at \sqs = 8\tev measured with the muon and the electron~\cite{LHCb-PAPER-2015-003} channels.
The combined result is
\begin{equation*}
\sigma_{\Z\to\ellp\ellm} = 94.9 \pm 0.2 \pm 0.6 \pm 1.1 \pm 1.1 \pb \, .
\end{equation*}

Uncertainties due to the \gec, the \lhc beam energy and the luminosity measurement are assumed to be fully correlated, while the other uncertainties are assumed to be uncorrelated.

\subsection{Ratios of cross-sections at {\boldmath \sqs} = 8 TeV}
\label{sec:wzcomb}

The ratios of electroweak boson production cross-sections are defined as
\begin{align}
\ratiow &= \frac{\cswp}{\cswm} \, , \\
\ratiowpz &= \frac{\cswp}{\csz} \, , \\
\ratiowmz &= \frac{\cswm}{\csz} \, , \\
\ratiowz &= \frac{\cswp+\cswm}{\csz} \, ,
\end{align}
and the muon charge asymmetry as a function of the muon pseudorapidity is defined as
\begin{align}
\asy(\eta_{i}) = \frac{\cswp(\eta_{i})-\cswm(\eta_{i})}{\cswp(\eta_{i})+\cswm(\eta_{i})} \, .
\end{align}

The sources of uncertainties contributing to the determination of the ratios are summarised in Table~\ref{tab:wzsyst}.
With respect to the systematic uncertainties on the cross-sections, many sources cancel or are reduced.
The luminosity uncertainty completely cancels in the ratios, as do the correlated components of the GEC efficiency uncertainty.
The trigger used to select both samples is identical and most of the uncertainty on the determination of the trigger efficiency cancels.
The uncertainties on the tracking and muon identification efficiencies partially cancel in the ratios of \W and \Z boson cross-sections, as do the uncertainties due to the proton beam energies.
The uncertainties on the purities of the \W and \Z boson selections are uncorrelated and the FSR uncertainties are taken to be uncorrelated.
The dominant uncertainties on the ratios are due to the purity and the size of the samples.
The correlation coefficients used in the uncertainty calculations are given in Tables~\ref{tab:corr1}--\ref{tab:corr7} in Appendix~\ref{sec:correlation}.

\begin{table}[!t]
\begin{center}
\caption{Summary of the relative uncertainties on the \ratiow, \ratiowpz, \ratiowmz and \ratiowz cross-section ratios.}
\label{tab:wzsyst}
\begin{tabular}{ccccc}
Source & \multicolumn{4}{c}{Uncertainty [\%]} \\ \hline
& \ratiow & \ratiowpz & \ratiowmz & \ratiowz \\ \hline
Statistical & $0.30$ & $0.33$ & $0.36$ & $0.31$ \\ \hline
Purity & $0.25$ & $0.35$ & $0.30$ & $0.30$ \\ 
Tracking  & $0.05$ & $0.22$ & $0.24$ & $0.23$ \\
Identification & $0.01$ & $0.11$ & $0.11$  & $0.11$ \\
Trigger & $0.04$ & $0.10$ & $0.09$ & $0.09$ \\
GEC & $0.13$ & $0.22$ & $0.23$ & $0.21$ \\
Selection  & $0.10$ & $0.24$ & $0.24$ & $0.23$ \\
Acceptance and FSR & $0.21$ & $0.21$ & $0.19$ & $0.17$  \\ \hline
Systematic & $0.37$ & $0.59$ & $0.56$ & $0.54$ \\ \hline
Beam energy & $0.14$ & $0.15$ & $0.29$ & $0.21$ \\ \hline
Total & $0.50$ & $0.69$ & $0.73$ & $0.66$ \\
\end{tabular}
\end{center}
\end{table}

\begin{figure}[!t]
\begin{center}
\includegraphics[width=0.8\textwidth]{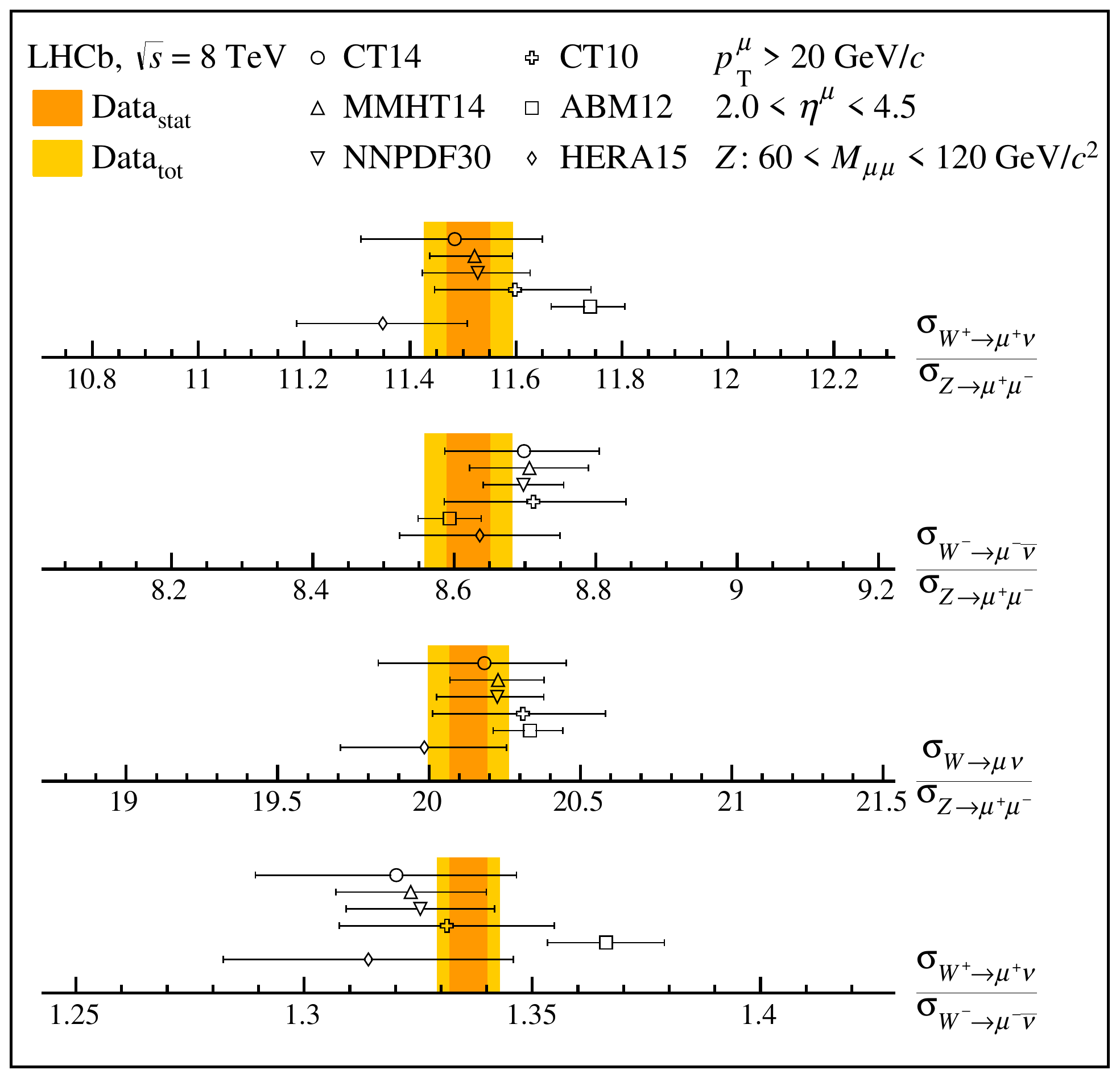}
\caption{Summary of the cross-section ratios. Measurements, represented as bands, are compared to (markers) \nnlo predictions with different parameterisations of the PDFs.}
\label{fig:summary_csr}
\end{center}
\end{figure}

\begin{figure}[!t]
\begin{center}
\includegraphics[width=0.84\textwidth]{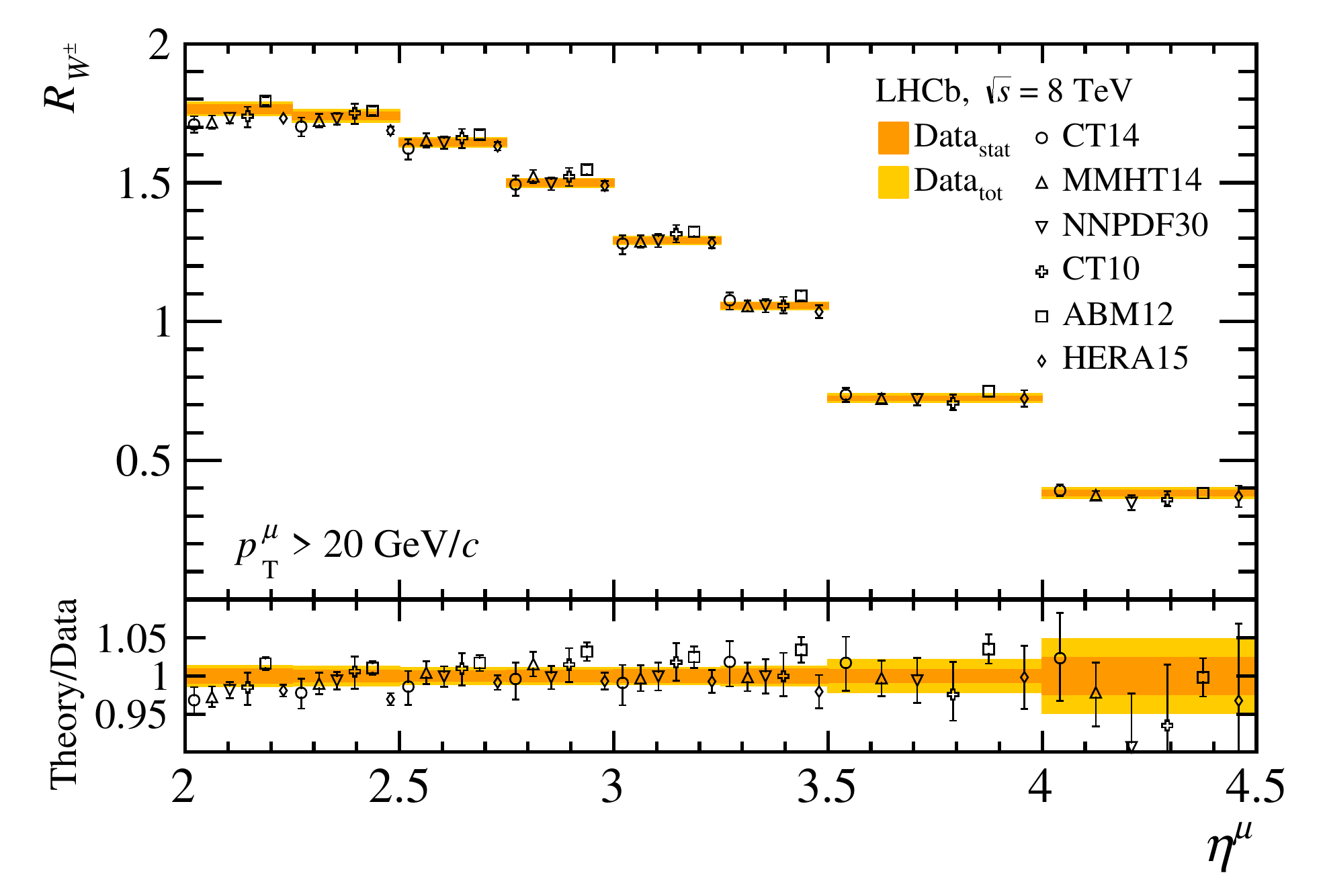}
\includegraphics[width=0.84\textwidth]{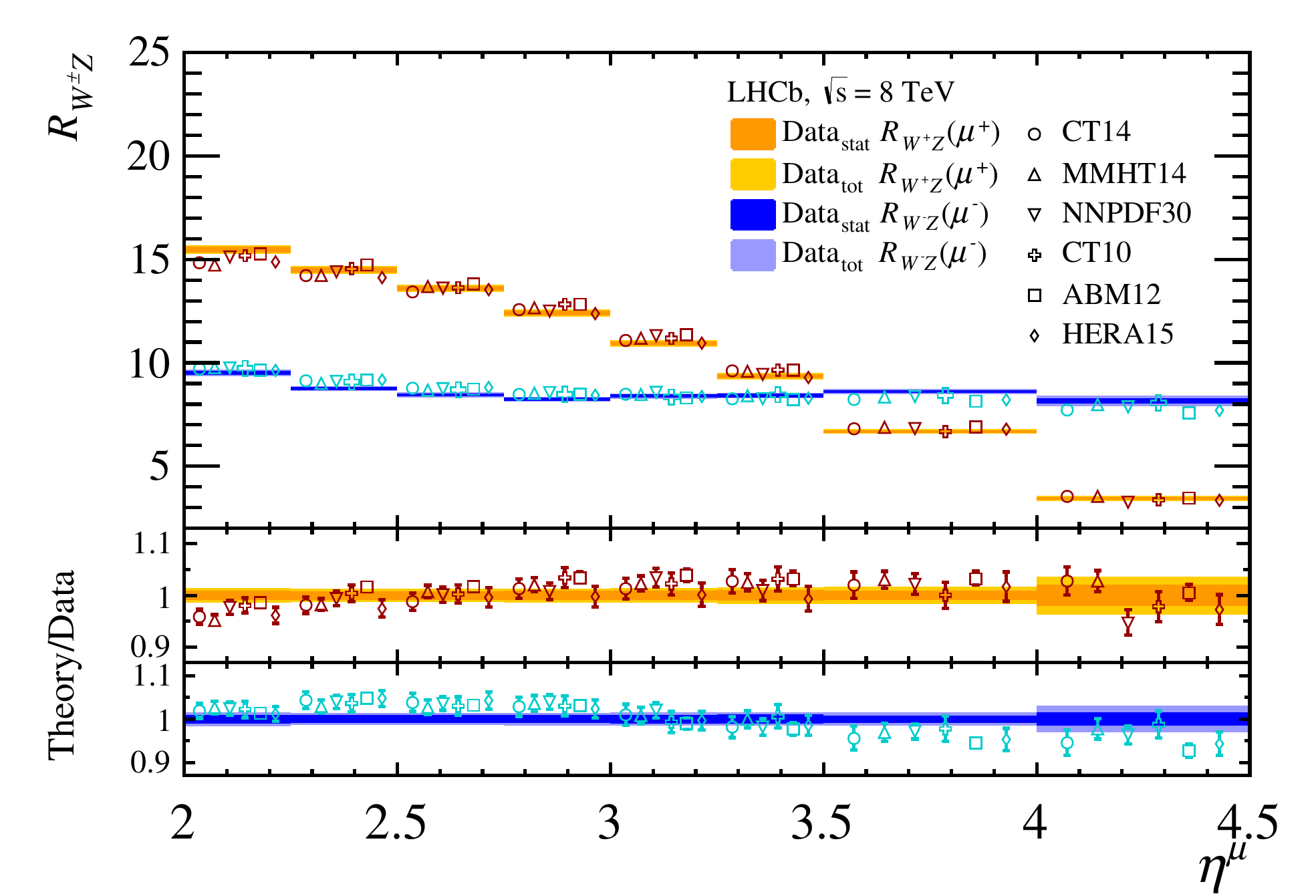}
\caption{(top) \Wp to \Wm cross-section ratio in bins of muon pseudorapidity. (bottom) \Wp (\Wm) to \Z cross-section ratio in bins of \mup (\mun) pseudorapidity. Measurements, represented as bands, are compared to (markers, displaced horizontally for presentation) NNLO predictions with different parameterisations of the PDFs.}
\label{fig:cswzr}
\end{center}
\end{figure}

\begin{figure}[!t]
\begin{center}
\includegraphics[width=0.84\textwidth]{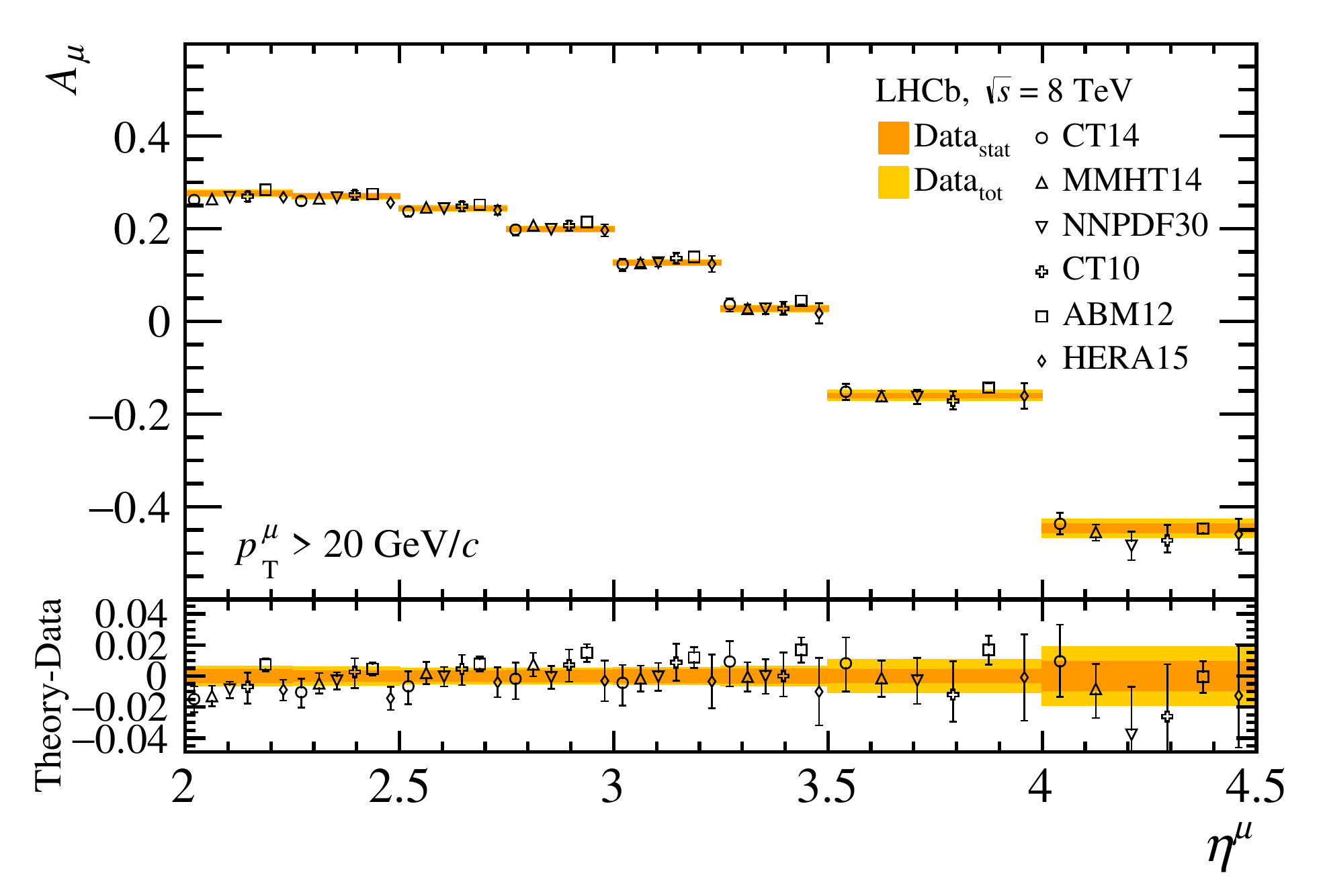}
\caption{\W production charge asymmetry in bins of muon pseudorapidity. Measurements, represented as bands, are compared to (markers, displaced horizontally for presentation) NNLO predictions with different parameterisations of the PDFs.}
\label{fig:cswa}
\end{center}
\end{figure}

The \W boson cross-section ratio is measured as
$$\ratiow = 1.336 \pm 0.004 \pm 0.005 \pm 0.002 \, ,$$
\noindent where the first uncertainty is statistical, the second is systematic and the third is due to the knowledge of the \lhc beam energy.
The \W to \Z boson production ratios are found to be
\begin{align*}
\ratiowpz &= 11.51 \pm 0.04 \pm 0.07 \pm 0.02 \, , \\
\ratiowmz &= \phantom{0}8.62 \pm 0.03 \pm 0.05 \pm 0.02 \, , \\
\ratiowz &= 20.13 \pm 0.06 \pm 0.11 \pm 0.04 \, .
\end{align*}
These measurements, as well as their predictions, are displayed in Fig.~\ref{fig:summary_csr}.
The data are well described by all PDF sets.
The \Wp to \Wm boson ratio, the charged \W to \Z boson ratios, and the muon charge asymmetry are determined differentially as a function of muon $\eta$, and displayed in Figs.~\ref{fig:cswzr} and \ref{fig:cswa}.
Good agreement between measured and predicted values is observed. 
All differential results are listed in Tables~\ref{tab:cswzr},~\ref{tab:cswr} and \ref{tab:cswa} of Appendix~\ref{sec:differential}.

\subsection{Ratios of cross-sections at different centre-of-mass energies}
\label{sec:rCOMsec}

The cross-section measurements detailed in the previous sections were also performed using 1.0\invfb of data at 7\tev~\cite{zmu}.
The two sets of measurements are used to make measurements of ratios of quantities at different centre-of-mass energies.
The ratios of cross-sections are defined as 
\begin{align}
R_{\Wp}^{8/7} &= \frac{\cswp^{8\tev}}{\cswp^{7\tev}} \, , \\
R_{\Wm}^{8/7} &= \frac{\cswm^{8\tev}}{\cswm^{7\tev}} \, , \\
R_{\Z}^{8/7} &= \frac{\csz^{8\tev}}{\csz^{7\tev}} \, ,
\end{align}
\noindent and the double ratios of cross-sections are defined as
\begin{align}
R_{\ratiow}^{8/7} &= \frac{\cswp^{8\tev}}{\cswp^{7\tev}}\frac{\cswm^{7\tev}}{\cswm^{8\tev}} \, , \\
R_{\ratiowpz}^{8/7} &= \frac{\cswp^{8\tev}}{\cswp^{7\tev}}\frac{\csz^{7\tev}}{\csz^{8\tev}} \, , \\
R_{\ratiowmz}^{8/7} &= \frac{\cswm^{8\tev}}{\cswm^{7\tev}}\frac{\csz^{7\tev}}{\csz^{8\tev}} \, , \\
R_{\ratiowz}^{8/7} &= \frac{\csw^{8\tev}}{\csw^{7\tev}}\frac{\csz^{7\tev}}{\csz^{8\tev}} \ . 
\end{align}

The following assumptions are made in order to estimate uncertainties on these ratios.
\begin{itemize}

\item The uncertainties due to statistically independent samples are uncorrelated, \eg the uncertainties due to the number of candidates in each measured bin, the uncertainties on the muon reconstruction efficiencies that are uncorrelated between \etamu bins, and the uncertainty that arises from corrections for having two muons inside the acceptance when measuring the selection efficiencies of \W bosons and the \W boson purity.

\item The uncertainties reflecting common methods are correlated, \eg the \Z candidate sample purity estimation, the components of the muon reconstruction efficiencies that are correlated between muon $\eta$ bins, and the uncertainty that arises when measuring selection efficiencies for \W bosons and all aspects of the \gec efficiency.

\item The uncertainty due to the \fsr correction is taken to be correlated in identical measurement bins and uncorrelated between different measurement bins. 

\item The beam energy has been directly measured for 4\tev beams with a precision of 0.65\%, but not for 3.5\tev beams~\cite{BEAM}. 
No additional uncertainty is expected to enter the energy measurement of 3.5\tev beams, so the relative uncertainty is taken to be the same, and fully correlated between data sets with 
different centre-of-mass energies.

\item The uncertainties ($\delta_{i}^{\sqrt{s}}$) entering the luminosity estimates are given in Ref.~\cite{lumi2}.
The degree of correlation between the luminosity measurements at different centre-of-mass energies is determined by assigning correlation coefficients ($c_{i}$) of 0, 1, [0,0.5], [0.5,1] or 
[0,1], where the last three represent intervals within which the true correlation is expected to lie. 
Pseudoexperiments are studied using correlation coefficients that are sampled from both uniform and arcsin distributions across these intervals. 
With this prescription, the total correlation is calculated using 
\begin{equation}
c = \frac{\sum_{i} c_{i} \, \delta_{i}^{8\tev} \delta_{i}^{7\tev} }{\delta^{8\tev} \delta^{7\tev}}
\end{equation}
and estimated to be $0.55 \pm 0.06$. 
A correlation coefficient of 0.55 is used.

\end{itemize}
A summary of the uncertainties on ratios of quantities at different centre-of-mass energies is given in Table~\ref{tab:wzCOMsyst}.

\begin{table}[!t]
\begin{center}
\caption{Summary of the relative uncertainties on the electroweak boson cross-section ratios at different centre-of-mass energies.}
\label{tab:wzCOMsyst}
\begin{tabular}{cccccccc}
Source & \multicolumn{7}{c}{Uncertainty [\%]} \\ \hline
	& $R_{\Wp}^{8/7}$ & $R_{\Wm}^{8/7}$ & $R_{\Z}^{8/7}$ & $R_{\ratiow}^{8/7}$ & $R_{\ratiowpz}^{8/7}$ & $R_{\ratiowmz}^{8/7}$ & $R_{\ratiowz}^{8/7}$ \\ \hline
    Statistical			& 0.30              & 0.37              & 0.49	& 0.48                   & 0.58                       & 0.62                       & 0.55            \\ \hline
    Purity		        & 0.41              & 0.45              & ---	& 0.65                   & 0.41                       & 0.45                       & 0.29            \\
    Tracking			& 0.33              & 0.27              & 0.53	& 0.09                   & 0.23                       & 0.26                       & 0.24            \\
    Identification		& 0.07              & 0.07              & 0.13	& 0.03                   & 0.07                       & 0.06                       & 0.07            \\
    Trigger			& 0.27              & 0.25              & 0.09	& 0.08                   & 0.19                       & 0.16                       & 0.17            \\
    GEC				& 0.15              & 0.14              & 0.09	& 0.07                   & 0.09                       & 0.09                       & 0.08            \\
    Selection			& 0.17              & 0.17              & ---	& 0.04                   & 0.17                       & 0.17                       & 0.16            \\
    Acceptance and FSR	        & 0.05              & 0.06              & 0.04	& 0.08                   & 0.07                       & 0.07                       & 0.06            \\ \hline
    Systematic			& 0.64              & 0.63              & 0.56	& 0.66                   & 0.55                       & 0.59                       & 0.46            \\ \hline
    Beam energy		        & 0.06              & 0.05              & 0.10	& ---                    & 0.04                       & 0.05                       & 0.05           \\ 
    Luminosity			& 1.45              & 1.45              & 1.45	& ---                    & ---                        & ---                        & ---            \\ \hline
    Total			& 1.61              & 1.62              & 1.63	& 0.82                   & 0.80                       & 0.86                       & 0.72            \\
\end{tabular}
\end{center}
\end{table}

\begin{figure}[!t]
\begin{center}
\includegraphics[width=0.8\textwidth]{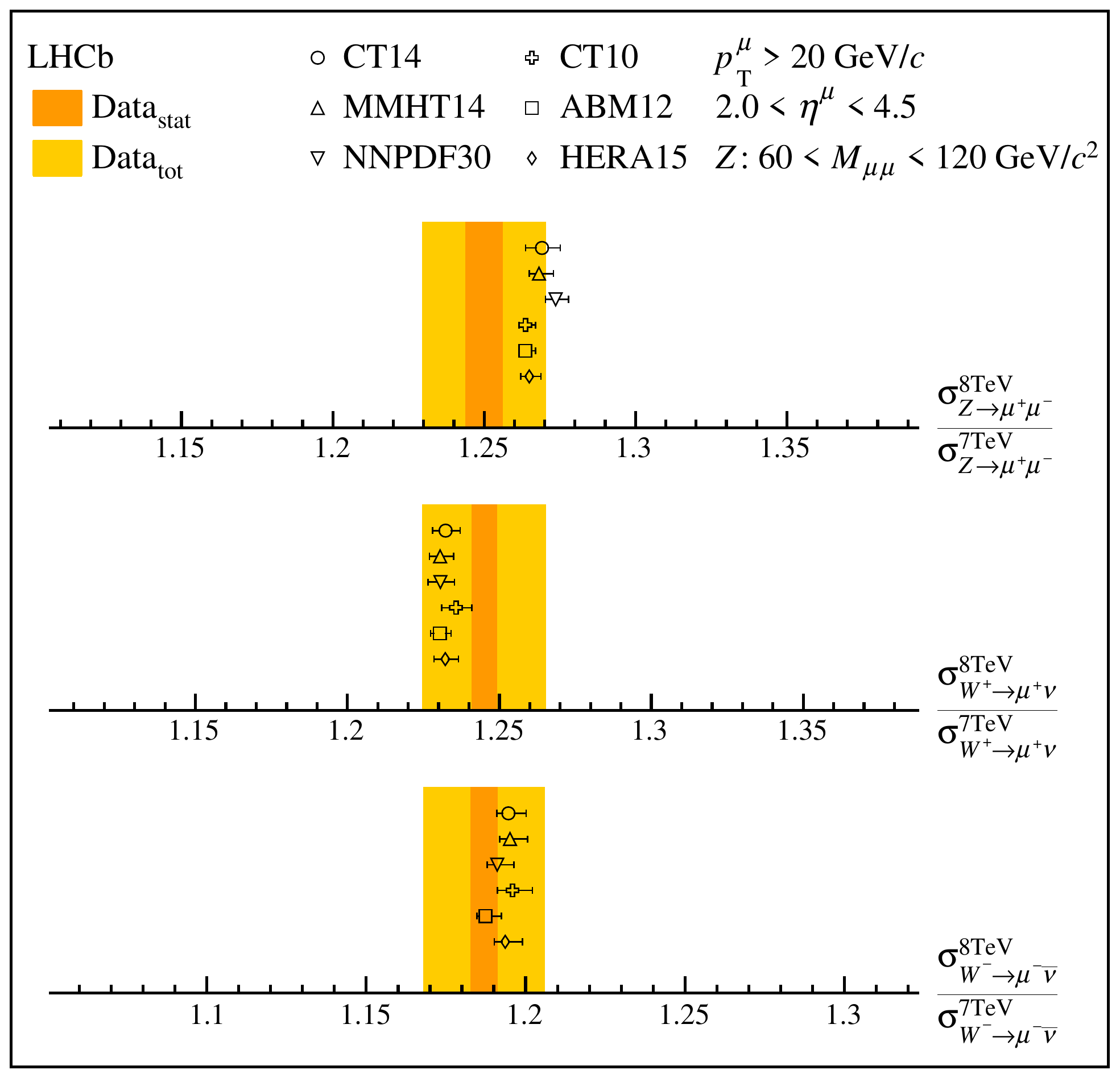}
\caption{Summary of the \W and \Z cross-section ratios at different centre-of-mass energies. Measurements, represented as bands, are compared to (markers) NNLO predictions with different parameterisations of the PDFs.}
\label{fig:summary_cs_com}
\end{center}
\end{figure}

The cross-section ratios at different centre-of-mass energies, measured for the same kinematic range as the total cross-sections, are
\begin{align*}
R_{\Wp}^{8/7} &= 1.245 \pm 0.004 \pm 0.008 \pm 0.001 \pm 0.018 \, , \\
R_{\Wm}^{8/7} &= 1.187 \pm 0.004 \pm 0.007 \pm 0.001 \pm 0.017 \, , \\
R_{\Z}^{8/7} &= 1.250 \pm 0.006 \pm 0.007 \pm 0.001 \pm 0.018 \, ,
\end{align*}
\noindent where the first uncertainties are statistical, the second are systematic, the third are due to the knowledge of the \lhc beam energy and the fourth are due to the luminosity measurement.
The measurements and predictions are in agreement, as shown in Fig.~\ref{fig:summary_cs_com}.
Compared to Fig.~\ref{fig:summary_cs}, the variation in the predictions is small.
This indicates that the uncertainty due to the PDF is very much reduced, which is also reflected in the calculated uncertainties on the individual PDF predictions.

\begin{figure}[!t]
\begin{center}
\includegraphics[width=0.8\textwidth]{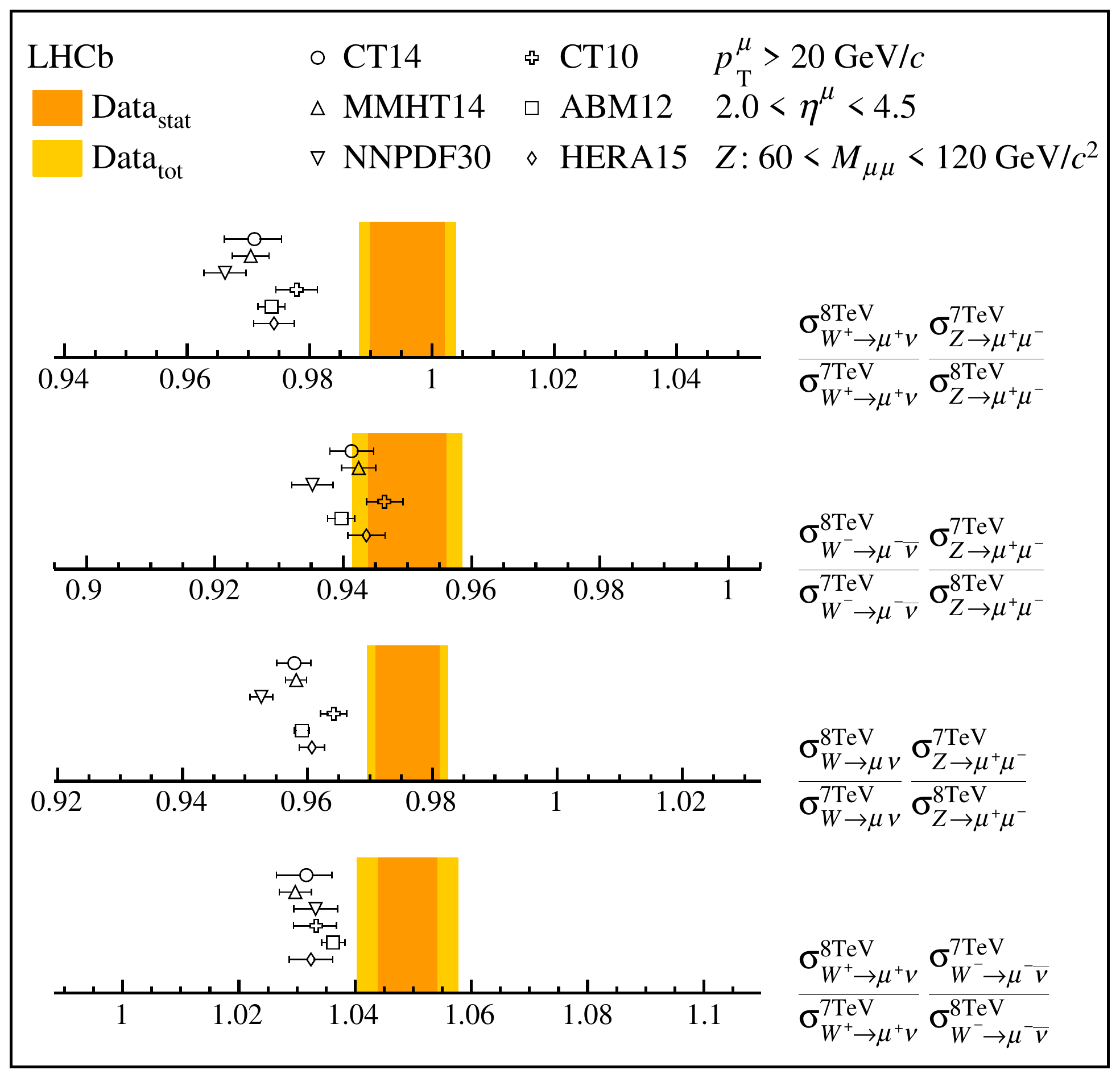}
\caption{Summary of the cross-section double ratios at different centre-of-mass energies. Measurements, represented as bands, are compared to (markers) NNLO predictions with different parameterisations of the PDFs.}
\label{fig:summary_csr_com}
\end{center}
\end{figure}

Even more precise tests can be obtained through the following double ratios of cross-sections, which are independent of the luminosities of either data set.
These double ratios are defined and measured as
\begin{align*}
R_{\ratiow}^{8/7}  &= 1.049 \pm 0.005 \pm 0.007 \, , \\
R_{\ratiowpz}^{8/7} &= 0.996 \pm 0.006 \pm 0.005 \, , \\
R_{\ratiowmz}^{8/7} &= 0.950 \pm 0.006 \pm 0.006 \, , \\
R_{\ratiowz}^{8/7} &= 0.976 \pm 0.005 \pm 0.004 \, ,
\end{align*}
\noindent where the first uncertainties are statistical and the second are systematic.
The largest source of systematic uncertainty on these ratios is due to the evaluation of the purity of the \W boson sample, which ranges between 0.3\% and 0.7\%.
The double ratios are shown in Fig.~\ref{fig:summary_csr_com}, where the uncertainties on the predictions due to the PDF, scale, \as and numerical integration are of similar magnitude.
Taking the uncertainty on the SM prediction to be reflected by the spread of the PDF predictions, the maximal deviation between the measured results and the theory is at the level of about 2 standard deviations.

\begin{figure}[!t]
\begin{center}
\includegraphics[width=0.75\textwidth]{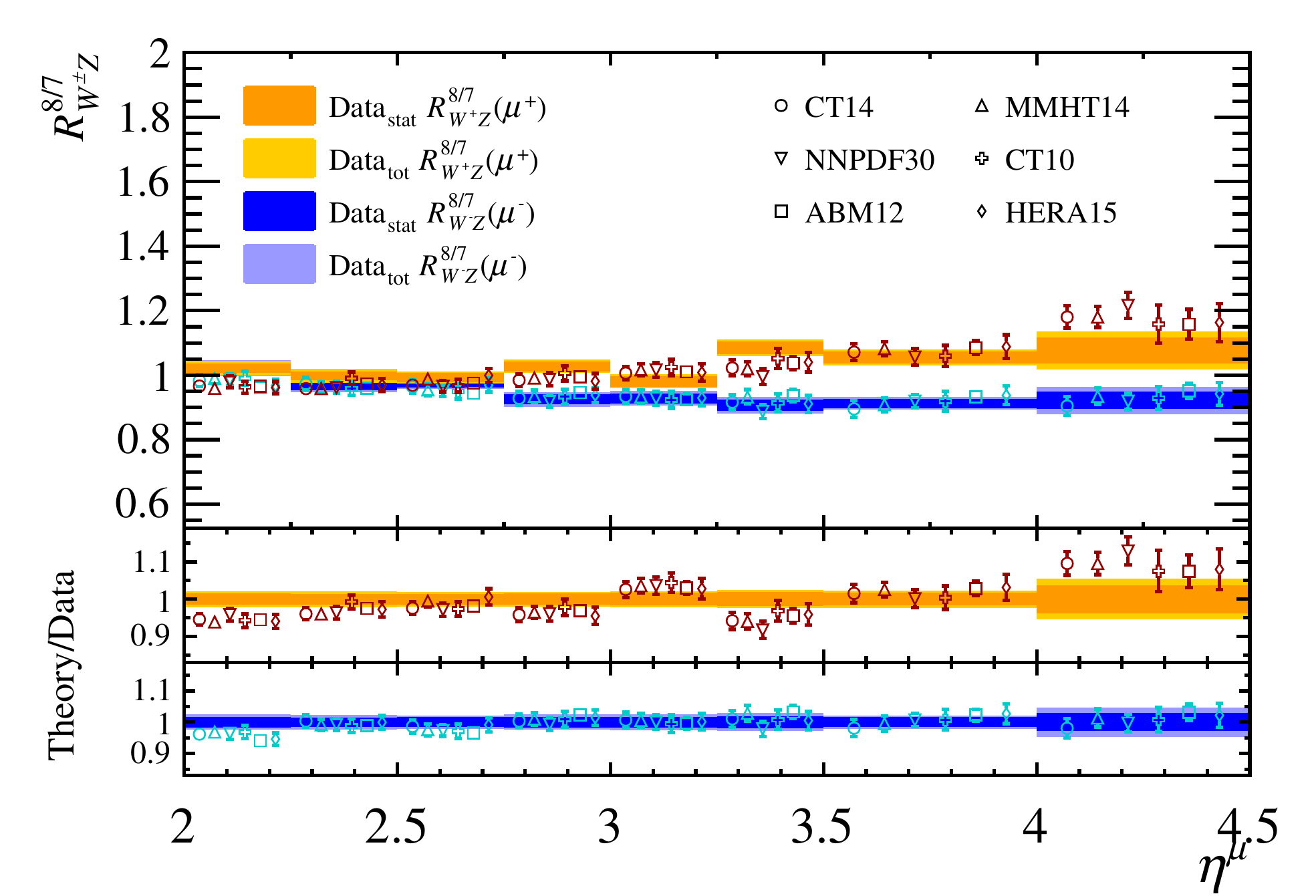}
\caption{Double ratios of cross-sections at different centre-of-mass energies as a function of muon pseudorapidity. Measurements, represented as bands, are compared to (markers, displaced horizontally for presentation) NNLO predictions with different parameterisations of the PDFs.}
\label{fig:summaryreta_c}
\end{center}
\end{figure}

The ratios $R_{\ratiowpz}^{8/7}$, $R_{\ratiowmz}^{8/7}$ are also measured differentially as a function of muon $\eta$.
These measurements are displayed in Fig.~\ref{fig:summaryreta_c}, where only uncertainties due to PDFs are included on the predictions. 
Good agreement between measurement and prediction is observed, especially for the $R_{\ratiowmz}^{8/7}$ ratio.
The $R_{\ratiowpz}^{8/7}$ ratio increases as a function of \etamu, while the $R_{\ratiowmz}^{8/7}$ ratio decreases as a function of \etamu. 
The PDF uncertainties are largest for the $R_{\ratiowpz}^{8/7}$ ratio at high pseudorapidity, suggesting that these measurements can improve the determination of the PDFs in this region.
Differential measurements are reported in Table~\ref{tab:cswzr87} of Appendix~\ref{sec:differential}, along with new differential measurements that were not published in Ref.~\cite{zmu} that 
are required for this analysis (Tables~\ref{tab:csZETA7} and \ref{tab:cswzr7}).

\section{Conclusions} 
\label{sec:conclusions}

Measurements of forward electroweak boson production at \sqs = 8\tev are presented and found to be in agreement with NNLO 
calculations in perturbative quantum chromodynamics. 
The large degree of correlation between the uncertainties allows for sub-percent determination of the cross-section ratios.
These represent the most precise determinations to date of electroweak boson production at the \lhc.
Using previous results from the \mbox{\sqs = 7\tev} data set, the evolution with the centre-of-mass energy is studied. 
Good agreement between measured and predicted cross-section ratios is observed. 
The experimental uncertainties are dominated by luminosity uncertainties of about 1.5\%.
Double ratios of cross-sections at different centre-of-mass energies are independent of the luminosity and are thus a more precise class of observables, determined with precision of between 0.7\% and 0.9\%.
In the cross-section ratios, the predictions deviate slightly from the measurements. 
Such deviations can be expected in BSM extensions that feature new sources of \W and \Z production. 
This motivates the extension of this analysis to higher energies, as will be possible with future data.

\section*{Acknowledgements}

\noindent We express our gratitude to our colleagues in the CERN
accelerator departments for the excellent performance of the LHC. We
thank the technical and administrative staff at the LHCb
institutes. We acknowledge support from CERN and from the national
agencies: CAPES, CNPq, FAPERJ and FINEP (Brazil); NSFC (China);
CNRS/IN2P3 (France); BMBF, DFG and MPG (Germany); INFN (Italy); 
FOM and NWO (The Netherlands); MNiSW and NCN (Poland); MEN/IFA (Romania); 
MinES and FANO (Russia); MinECo (Spain); SNSF and SER (Switzerland); 
NASU (Ukraine); STFC (United Kingdom); NSF (USA).
We acknowledge the computing resources that are provided by CERN, IN2P3 (France), KIT and DESY (Germany), INFN (Italy), SURF (The Netherlands), PIC (Spain), GridPP (United Kingdom), RRCKI (Russia), CSCS (Switzerland), IFIN-HH (Romania), CBPF (Brazil), PL-GRID (Poland) and OSC (USA). We are indebted to the communities behind the multiple open 
source software packages on which we depend. We are also thankful for the 
computing resources and the access to software R\&D tools provided by Yandex LLC (Russia).
Individual groups or members have received support from AvH Foundation (Germany),
EPLANET, Marie Sk\l{}odowska-Curie Actions and ERC (European Union), 
Conseil G\'{e}n\'{e}ral de Haute-Savoie, Labex ENIGMASS and OCEVU, 
R\'{e}gion Auvergne (France), RFBR (Russia), GVA, XuntaGal and GENCAT (Spain), The Royal Society and Royal Commission for the Exhibition of 1851 (United Kingdom).

\clearpage

{\noindent\bf\Large Appendices}

\appendix

\section{Differential measurements}
\label{sec:differential}

Differential production cross-section measurements as functions of the pseudorapidities of the muons from the decay of the \Z 
boson were not included in the previous publication that describes the analysis of the $\sqrt{s}$ = 7 TeV data 
set~\cite{zmu}.
They are provided in this appendix in Table~\ref{tab:csZETA7}, along with the related measurements of the differential \W to 
\Z boson cross-section ratios in Table~\ref{tab:cswzr7}.

\begin{table}[b]
\begin{center}
\caption{Cross-section for (top) \Wp and (bottom) \Wm boson production in bins of muon pseudorapidity. The first uncertainties are statistical, the second are systematic, the third are due to the knowledge of the \lhc beam energy and the fourth are due to the luminosity measurement. The last column lists the final-state radiation correction.}
\label{tab:csWETA}
\begin{tabular}{ccc}
\etamu & \cswp [\pb] & $\ffsr^{\Wp}$ \\ \hline
2.00 -- 2.25 & $236.5 \pm 1.2 \pm 3.2 \pm 2.4 \pm 2.7$           & $1.0188 \pm 0.0047$ \\
2.25 -- 2.50 & $208.4 \pm 0.9 \pm 2.2 \pm 2.1 \pm 2.4$           & $1.0163 \pm 0.0028$ \\
2.50 -- 2.75 & $182.0 \pm 0.8 \pm 1.8 \pm 1.8 \pm 2.1$           & $1.0158 \pm 0.0025$ \\
2.75 -- 3.00 & $153.3 \pm 0.7 \pm 1.6 \pm 1.5 \pm 1.8$           & $1.0148 \pm 0.0028$ \\
3.00 -- 3.25 & $119.5 \pm 0.6 \pm 1.3 \pm 1.2 \pm 1.4$           & $1.0152 \pm 0.0032$ \\
3.25 -- 3.50 & $\phantom{0}84.4 \pm 0.5 \pm 1.0 \pm 0.8 \pm 1.0$ & $1.0150 \pm 0.0046$ \\
3.50 -- 4.00 & $\phantom{0}86.4 \pm 0.5 \pm 1.2 \pm 0.9 \pm 1.0$ & $1.0175 \pm 0.0045$ \\
4.00 -- 4.50 & $\phantom{0}23.0 \pm 0.4 \pm 0.7 \pm 0.2 \pm 0.3$ & $1.0211 \pm 0.0087$ \\
\end{tabular}

\vspace{0.5cm}

\begin{tabular}{ccc}
\etamu & \cswm [\pb] & $\ffsr^{\Wm}$ \\ \hline
2.00 -- 2.25 & $134.0 \pm 0.9 \pm 1.8 \pm 1.2 \pm 1.6$           & $1.0172 \pm 0.0026$ \\
2.25 -- 2.50 & $119.8 \pm 0.7 \pm 1.4 \pm 1.0 \pm 1.4$           & $1.0155 \pm 0.0027$ \\
2.50 -- 2.75 & $110.6 \pm 0.6 \pm 1.2 \pm 1.0 \pm 1.3$           & $1.0153 \pm 0.0028$ \\
2.75 -- 3.00 & $102.4 \pm 0.6 \pm 1.2 \pm 0.9 \pm 1.2$           & $1.0162 \pm 0.0030$ \\
3.00 -- 3.25 & $\phantom{0}92.5 \pm 0.6 \pm 1.1 \pm 0.8 \pm 1.1$ & $1.0160 \pm 0.0031$ \\
3.25 -- 3.50 & $\phantom{0}79.9 \pm 0.5 \pm 0.9 \pm 0.7 \pm 0.9$ & $1.0176 \pm 0.0033$ \\
3.50 -- 4.00 & $119.3 \pm 0.6 \pm 1.5 \pm 1.0 \pm 1.4$           & $1.0200 \pm 0.0033$ \\
4.00 -- 4.50 & $\phantom{0}60.0 \pm 0.7 \pm 1.6 \pm 0.5 \pm 0.7$ & $1.0243 \pm 0.0053$ \\
\end{tabular}
\end{center}
\end{table}

\begin{table}[h]
\begin{center}
\caption{Cross-section for \Z boson production in bins of boson rapidity. The first uncertainties are statistical, the second are systematic, the third are due to the knowledge of the \lhc beam energy and the fourth are due to the luminosity measurement. The last column lists the final-state radiation correction.}
\label{tab:csZY}
\begin{tabular}{ccc}
\rapz & \csz [\pb] & $\ffsr^{\Z}$ \\ \hline
2.000 -- 2.125 &     1.223        $\pm$     0.033 $\pm$ 0.055 $\pm$ 0.014 $\pm$ 0.014 & $1.0466 \pm 0.0395$ \\
2.125 -- 2.250 &     3.263        $\pm$     0.051 $\pm$ 0.060 $\pm$ 0.038 $\pm$ 0.038 & $1.0305 \pm 0.0119$ \\
2.250 -- 2.375 &     4.983        $\pm$     0.062 $\pm$ 0.064 $\pm$ 0.057 $\pm$ 0.058 & $1.0277 \pm 0.0069$ \\
2.375 -- 2.500 &     6.719        $\pm$     0.070 $\pm$ 0.072 $\pm$ 0.077 $\pm$ 0.078 & $1.0252 \pm 0.0061$ \\
2.500 -- 2.625 &     8.051        $\pm$     0.076 $\pm$ 0.074 $\pm$ 0.093 $\pm$ 0.094 & $1.0264 \pm 0.0048$ \\
2.625 -- 2.750 &     8.967        $\pm$     0.079 $\pm$ 0.074 $\pm$ 0.103 $\pm$ 0.105 & $1.0257 \pm 0.0032$ \\
2.750 -- 2.875 &     9.561        $\pm$     0.081 $\pm$ 0.076 $\pm$ 0.110 $\pm$ 0.112 & $1.0258 \pm 0.0038$ \\
2.875 -- 3.000 &     9.822        $\pm$     0.082 $\pm$ 0.071 $\pm$ 0.113 $\pm$ 0.115 & $1.0252 \pm 0.0027$ \\
3.000 -- 3.125 &     9.721        $\pm$     0.081 $\pm$ 0.074 $\pm$ 0.112 $\pm$ 0.114 & $1.0282 \pm 0.0035$ \\
3.125 -- 3.250 &     9.030        $\pm$     0.078 $\pm$ 0.071 $\pm$ 0.104 $\pm$ 0.105 & $1.0264 \pm 0.0030$ \\
3.250 -- 3.375 &     7.748        $\pm$     0.072 $\pm$ 0.074 $\pm$ 0.089 $\pm$ 0.090 & $1.0261 \pm 0.0066$ \\
3.375 -- 3.500 &     6.059        $\pm$     0.063 $\pm$ 0.051 $\pm$ 0.070 $\pm$ 0.071 & $1.0248 \pm 0.0040$ \\
3.500 -- 3.625 &     4.385        $\pm$     0.054 $\pm$ 0.041 $\pm$ 0.050 $\pm$ 0.051 & $1.0258 \pm 0.0060$ \\
3.625 -- 3.750 &     2.724        $\pm$     0.042 $\pm$ 0.027 $\pm$ 0.031 $\pm$ 0.032 & $1.0228 \pm 0.0053$ \\
3.750 -- 3.875 &     1.584        $\pm$     0.032 $\pm$ 0.020 $\pm$ 0.018 $\pm$ 0.019 & $1.0180 \pm 0.0079$ \\
3.875 -- 4.000 &     0.749        $\pm$     0.022 $\pm$ 0.012 $\pm$ 0.009 $\pm$ 0.009 & $1.0207 \pm 0.0100$ \\
4.000 -- 4.250 &     0.383        $\pm$     0.016 $\pm$ 0.008 $\pm$ 0.004 $\pm$ 0.004 & $1.0183 \pm 0.0140$ \\
4.250 -- 4.500 &     0.011        $\pm$     0.003 $\pm$ 0.001 $\pm$ 0.000 $\pm$ 0.000 & $1.0177 \pm 0.0761$ \\
\end{tabular}
\end{center}
\end{table}

\begin{table}[h]
\begin{center}
\caption{Cross-section for \Z boson production in bins of boson transverse momentum. The first uncertainties are statistical, the second are systematic, the third are due to the knowledge of the \lhc beam energy and the fourth are due to the luminosity measurement. The last column lists the final-state radiation correction.}
\label{tab:csZPT}
\begin{tabular}{ccc}  
\ptz [\gevc] & \csz [\pb] & $\ffsr^{\Z}$ \\ \hline  
\phantom{0}0.0 -- \phantom{00}2.2 & 7.903           $\pm$     0.082$\pm$ 0.130 $\pm$ 0.091 $\pm$ 0.092    & $1.0962 \pm 0.0045$ \\ 
\phantom{0}2.2 -- \phantom{00}3.4 & 7.705           $\pm$     0.080$\pm$ 0.108 $\pm$ 0.089 $\pm$ 0.090    & $1.0788 \pm 0.0055$ \\ 
\phantom{0}3.4 -- \phantom{00}4.6 & 7.609           $\pm$     0.078$\pm$ 0.080 $\pm$ 0.088 $\pm$ 0.089    & $1.0620 \pm 0.0039$ \\ 
\phantom{0}4.6 -- \phantom{00}5.8 & 7.073           $\pm$     0.075$\pm$ 0.078 $\pm$ 0.081 $\pm$ 0.083    & $1.0472 \pm 0.0035$ \\ 
\phantom{0}5.8 -- \phantom{00}7.2 & 7.379           $\pm$     0.078$\pm$ 0.069 $\pm$ 0.085 $\pm$ 0.086    & $1.0290 \pm 0.0044$ \\ 
\phantom{0}7.2 -- \phantom{00}8.7 & 6.813           $\pm$     0.076$\pm$ 0.074 $\pm$ 0.078 $\pm$ 0.080    & $1.0165 \pm 0.0060$ \\ 
\phantom{0}8.7 -- \phantom{0}10.5 & 6.751           $\pm$     0.075$\pm$ 0.064 $\pm$ 0.078 $\pm$ 0.079   & $1.0044 \pm 0.0037$ \\ 
10.5 -- \phantom{0}12.8 & 7.204           $\pm$     0.078$\pm$ 0.073 $\pm$ 0.083 $\pm$ 0.084  & $0.9953 \pm 0.0060$ \\ 
12.8 -- \phantom{0}15.4 & 6.270           $\pm$     0.073$\pm$ 0.053 $\pm$ 0.072 $\pm$ 0.073  & $0.9852 \pm 0.0035$ \\ 
15.4 -- \phantom{0}19.0 & 6.534           $\pm$     0.072$\pm$ 0.064 $\pm$ 0.075 $\pm$ 0.076  & $0.9830 \pm 0.0042$ \\ 
19.0 -- \phantom{0}24.5 & 6.953           $\pm$     0.071$\pm$ 0.066 $\pm$ 0.080 $\pm$ 0.081  & $0.9853 \pm 0.0044$ \\ 
24.5 -- \phantom{0}34.0 & 6.999           $\pm$     0.069$\pm$ 0.062 $\pm$ 0.080 $\pm$ 0.082  & $1.0109 \pm 0.0031$ \\ 
34.0 -- \phantom{0}63.0 & 7.602           $\pm$     0.070$\pm$ 0.072 $\pm$ 0.087 $\pm$ 0.089  & $1.0380 \pm 0.0034$ \\ 
63.0 -- 270.0 & 2.176           $\pm$     0.037$\pm$ 0.025 $\pm$ 0.025 $\pm$ 0.025 & $1.0604 \pm 0.0059$ \\ 
\end{tabular}
\end{center}
\end{table}

\begin{table}[h]
\begin{center}
\caption{Cross-section for \Z boson production in bins of boson \phist. The first uncertainties are statistical, the second are systematic, the third are due to the knowledge of the \lhc beam energy and the fourth are due to the luminosity measurement. The last column lists the final-state radiation correction.}
\label{tab:csZPHI}
\begin{tabular}{ccc}
\phist & \csz [\pb] & $\ffsr^{\Z}$ \\ \hline
0.00 -- 0.01 & 10.442           $\pm$ 0.077 $\pm$ 0.118 $\pm$ 0.120 $\pm$ 0.122           & $1.0367 \pm 0.0028$ \\  
0.01 -- 0.02 & \phantom{0}9.704           $\pm$ 0.076 $\pm$ 0.116 $\pm$ 0.112 $\pm$ 0.113 & $1.0346 \pm 0.0031$ \\  
0.02 -- 0.03 & \phantom{0}8.510           $\pm$ 0.071 $\pm$ 0.130 $\pm$ 0.098 $\pm$ 0.099 & $1.0323 \pm 0.0031$ \\  
0.03 -- 0.05 & 13.749           $\pm$ 0.089 $\pm$ 0.151 $\pm$ 0.158 $\pm$ 0.161           & $1.0288 \pm 0.0024$ \\  
0.05 -- 0.07 & 10.085           $\pm$ 0.076 $\pm$ 0.119 $\pm$ 0.116 $\pm$ 0.118           & $1.0254 \pm 0.0036$ \\  
0.07 -- 0.10 & 10.662           $\pm$ 0.077 $\pm$ 0.159 $\pm$ 0.123 $\pm$ 0.125           & $1.0211 \pm 0.0030$ \\  
0.10 -- 0.15 & 10.575           $\pm$ 0.077 $\pm$ 0.133 $\pm$ 0.122 $\pm$ 0.123           & $1.0196 \pm 0.0029$ \\  
0.15 -- 0.20 & \phantom{0}6.322           $\pm$ 0.059 $\pm$ 0.074 $\pm$ 0.073 $\pm$ 0.074 & $1.0177 \pm 0.0034$ \\  
0.20 -- 0.30 & \phantom{0}6.681           $\pm$ 0.061 $\pm$ 0.085 $\pm$ 0.077 $\pm$ 0.078 & $1.0188 \pm 0.0039$ \\  
0.30 -- 0.40 & \phantom{0}3.213           $\pm$ 0.042 $\pm$ 0.064 $\pm$ 0.037 $\pm$ 0.038 & $1.0210 \pm 0.0064$ \\  
0.40 -- 0.60 & \phantom{0}2.837           $\pm$ 0.040 $\pm$ 0.055 $\pm$ 0.033 $\pm$ 0.033 & $1.0251 \pm 0.0065$ \\  
0.60 -- 0.80 & \phantom{0}1.030           $\pm$ 0.024 $\pm$ 0.027 $\pm$ 0.012 $\pm$ 0.012 & $1.0258 \pm 0.0114$ \\  
0.80 -- 1.20 & \phantom{0}0.670           $\pm$ 0.020 $\pm$ 0.030 $\pm$ 0.008 $\pm$ 0.008 & $1.0269 \pm 0.0110$ \\  
1.20 -- 2.00 & \phantom{0}0.263           $\pm$ 0.013 $\pm$ 0.022 $\pm$ 0.003 $\pm$ 0.003 & $1.0276 \pm 0.0210$ \\  
2.00 -- 4.00 & \phantom{0}0.094           $\pm$ 0.008 $\pm$ 0.023 $\pm$ 0.001 $\pm$ 0.001 & $1.0345 \pm 0.0396$ \\  
\end{tabular}
\end{center}
\end{table}

\begin{table}[h]
\begin{center}
\caption{Cross-section for \Z boson production in bins of muon pseudorapidity. The first uncertainties are statistical, the second are systematic, the third are due to the knowledge of the \lhc beam energy and the fourth are due to the luminosity measurement. The last column lists the final-state radiation correction.}
\label{tab:csZETA}
\begin{tabular}{cccc}
\etamu & $\csz(\etamup)$ [\pb] & $\ffsr^{\Z}$ \\ \hline
2.00 -- 2.25 &     15.28        $\pm$     0.11 $\pm$ 0.18 $\pm$ 0.18 $\pm$ 0.18 &           $1.0293 \pm 0.0036$ \\
2.25 -- 2.50 &     14.39        $\pm$     0.10 $\pm$ 0.13 $\pm$ 0.17 $\pm$ 0.17 &           $1.0250 \pm 0.0027$ \\
2.50 -- 2.75 &     13.39        $\pm$     0.10 $\pm$ 0.11 $\pm$ 0.15 $\pm$ 0.16 &           $1.0244 \pm 0.0044$ \\
2.75 -- 3.00 &     12.37        $\pm$     0.09 $\pm$ 0.10 $\pm$ 0.14 $\pm$ 0.14 &           $1.0240 \pm 0.0033$ \\
3.00 -- 3.25 &     10.93        $\pm$     0.09 $\pm$ 0.09 $\pm$ 0.13 $\pm$ 0.13 &           $1.0234 \pm 0.0037$ \\
3.25 -- 3.50 &     \phantom{0}9.02        $\pm$     0.08 $\pm$ 0.08 $\pm$ 0.10 $\pm$ 0.11 & $1.0246 \pm 0.0046$ \\
3.50 -- 4.00 &     12.94        $\pm$     0.09 $\pm$ 0.10 $\pm$ 0.15 $\pm$ 0.15 &           $1.0269 \pm 0.0033$ \\
4.00 -- 4.50 &     \phantom{0}6.67        $\pm$     0.07 $\pm$ 0.07 $\pm$ 0.08 $\pm$ 0.08 & $1.0365 \pm 0.0038$ \\
\end{tabular}

\vspace{0.5cm}

\begin{tabular}{cccc}
\etamu & $\csz(\etamum)$ [\pb] & $\ffsr^{\Z}$ \\ \hline
2.00 -- 2.25 &     14.07        $\pm$     0.10 $\pm$ 0.18 $\pm$ 0.16 $\pm$ 0.16           &$1.0291 \pm 0.0056$ \\
2.25 -- 2.50 &     13.68        $\pm$     0.10 $\pm$ 0.13 $\pm$ 0.16 $\pm$ 0.16           &$1.0254 \pm 0.0028$ \\
2.50 -- 2.75 &     13.09        $\pm$     0.10 $\pm$ 0.10 $\pm$ 0.15 $\pm$ 0.15           &$1.0251 \pm 0.0028$ \\
2.75 -- 3.00 &     12.43        $\pm$     0.09 $\pm$ 0.11 $\pm$ 0.14 $\pm$ 0.15           &$1.0239 \pm 0.0033$ \\
3.00 -- 3.25 &     11.01        $\pm$     0.09 $\pm$ 0.10 $\pm$ 0.13 $\pm$ 0.13           &$1.0227 \pm 0.0041$ \\
3.25 -- 3.50 &     \phantom{0}9.49        $\pm$     0.08 $\pm$ 0.08 $\pm$ 0.11 $\pm$ 0.11 &$1.0246 \pm 0.0042$ \\
3.50 -- 4.00 &     13.85        $\pm$     0.10 $\pm$ 0.11 $\pm$ 0.16 $\pm$ 0.16           &$1.0268 \pm 0.0036$ \\
4.00 -- 4.50 &     \phantom{0}7.35        $\pm$     0.07 $\pm$ 0.07 $\pm$ 0.09 $\pm$ 0.09 &$1.0353 \pm 0.0055$ \\
\end{tabular}
\end{center}
\end{table}

\clearpage

\begin{table}[h]
\begin{center}
\caption{(Top) \Wp and (bottom) \Wm to \Z cross-section ratios in bins of muon pseudorapidity. The first uncertainties are statistical, the second are systematic, the third are due to the knowledge of the \lhc beam energy and the fourth are due to the luminosity measurement.}
\label{tab:cswzr}
\begin{tabular}{cc}
\etamup & \ratiowpz \\ \hline
2.00 -- 2.25 &     15.478        $\pm$     0.134 $\pm$ 0.174 $\pm$ 0.024 $\pm$ 0.001 \\
2.25 -- 2.50 &     14.490        $\pm$     0.119 $\pm$ 0.136 $\pm$ 0.022 $\pm$ 0.001 \\
2.50 -- 2.75 &     13.593        $\pm$     0.112 $\pm$ 0.137 $\pm$ 0.020 $\pm$ 0.001 \\
2.75 -- 3.00 &     12.406        $\pm$     0.108 $\pm$ 0.126 $\pm$ 0.019 $\pm$ 0.001 \\
3.00 -- 3.25 &     10.937        $\pm$     0.102 $\pm$ 0.115 $\pm$ 0.016 $\pm$ 0.001 \\
3.25 -- 3.50 &     \phantom{0}9.353        $\pm$     0.097 $\pm$ 0.114 $\pm$ 0.015 $\pm$ 0.001 \\
3.50 -- 4.00 &     \phantom{0}6.677        $\pm$     0.063 $\pm$ 0.093 $\pm$ 0.010 $\pm$ 0.001 \\
4.00 -- 4.50 &     \phantom{0}3.444        $\pm$     0.072 $\pm$ 0.103 $\pm$ 0.005 $\pm$ 0.000 \\
\end{tabular}
\end{center}

\begin{center}
\begin{tabular}{cc}
\etamum & \ratiowmz \\ \hline
2.00 -- 2.25 &     9.521        $\pm$     0.095 $\pm$ 0.117 $\pm$ 0.028 $\pm$ 0.001 \\
2.25 -- 2.50 &     8.754        $\pm$     0.080 $\pm$ 0.090 $\pm$ 0.025 $\pm$ 0.001 \\
2.50 -- 2.75 &     8.449        $\pm$     0.076 $\pm$ 0.086 $\pm$ 0.025 $\pm$ 0.001 \\
2.75 -- 3.00 &     8.235        $\pm$     0.077 $\pm$ 0.089 $\pm$ 0.024 $\pm$ 0.000 \\
3.00 -- 3.25 &     8.400        $\pm$     0.082 $\pm$ 0.096 $\pm$ 0.024 $\pm$ 0.001 \\
3.25 -- 3.50 &     8.414        $\pm$     0.088 $\pm$ 0.096 $\pm$ 0.024 $\pm$ 0.000 \\
3.50 -- 4.00 &     8.615        $\pm$     0.075 $\pm$ 0.107 $\pm$ 0.025 $\pm$ 0.001 \\
4.00 -- 4.50 &     8.166        $\pm$     0.130 $\pm$ 0.215 $\pm$ 0.023 $\pm$ 0.000 \\
\end{tabular}
\end{center}
\end{table}

\clearpage

\begin{table}[h]
\begin{center}
\caption{\Wp to \Wm cross-section ratio in bins of muon pseudorapidity. The first uncertainties are statistical, the second are systematic and the third are due to the knowledge of the \lhc beam energy.}
\label{tab:cswr}
\begin{tabular}{cc}
\etamu & \ratiow \\ \hline
2.00 -- 2.25 & $1.765 \pm 0.015 \pm 0.018 \pm 0.003$ \\
2.25 -- 2.50 & $1.740 \pm 0.012 \pm 0.018 \pm 0.002$ \\
2.50 -- 2.75 & $1.645 \pm 0.011 \pm 0.013 \pm 0.002$ \\
2.75 -- 3.00 & $1.499 \pm 0.011 \pm 0.011 \pm 0.002$ \\
3.00 -- 3.25 & $1.292 \pm 0.010 \pm 0.010 \pm 0.002$ \\
3.25 -- 3.50 & $1.057 \pm 0.009 \pm 0.009 \pm 0.002$ \\
3.50 -- 4.00 & $0.724 \pm 0.006 \pm 0.014 \pm 0.001$ \\
4.00 -- 4.50 & $0.383 \pm 0.009 \pm 0.016 \pm 0.001$ \\
\end{tabular}
\end{center}
\end{table}

\begin{table}[h]
\begin{center}
\caption{Lepton charge asymmetry in bins of muon pseudorapidity. The first uncertainties are statistical, the second are systematic and the third are due to the knowledge of the \lhc beam energy.}
\label{tab:cswa}
\begin{tabular}{cc}
\etamu & \asy (\%) \\ \hline
2.00 -- 2.25 & $\phantom{-}27.67 \pm 0.39 \pm 0.48 \pm 0.07$ \\
2.25 -- 2.50 & $\phantom{-}27.02 \pm 0.33 \pm 0.47 \pm 0.07$ \\
2.50 -- 2.75 & $\phantom{-}24.39 \pm 0.32 \pm 0.37 \pm 0.07$ \\
2.75 -- 3.00 & $\phantom{-}19.96 \pm 0.35 \pm 0.34 \pm 0.07$ \\
3.00 -- 3.25 & $\phantom{-}12.74 \pm 0.38 \pm 0.37 \pm 0.07$ \\
3.25 -- 3.50 & $\phantom{-0}2.75 \pm 0.43 \pm 0.42 \pm 0.07$ \\
3.50 -- 4.00 & $-15.99 \pm 0.39 \pm 0.96 \pm 0.07$ \\
4.00 -- 4.50 & $-44.63 \pm 0.89 \pm 1.64 \pm 0.06$ \\
\end{tabular}
\end{center}
\end{table}

\clearpage

\begin{table}[h]
\begin{center}
\caption{Ratios of (top) \Wp and (bottom) \Wm to \Z cross-section ratios at different centre-of-mass energies in bins of muon pseudorapidity. The first uncertainty is statistical and the second is systematic.}
\label{tab:cswzr87}
\begin{tabular}{cc}
\etamup & $R_{\ratiowpz}^{8/7}$ \\[0.1cm] \hline
2.00 -- 2.25 &     1.022        $\pm$     0.015 $\pm$ 0.016 \\
2.25 -- 2.50 &     0.997        $\pm$     0.014 $\pm$ 0.016 \\
2.50 -- 2.75 &     0.993        $\pm$     0.014 $\pm$ 0.013 \\
2.75 -- 3.00 &     1.027        $\pm$     0.016 $\pm$ 0.013 \\
3.00 -- 3.25 &     0.981        $\pm$     0.017 $\pm$ 0.014 \\
3.25 -- 3.50 &     1.085        $\pm$     0.020 $\pm$ 0.017 \\
3.50 -- 4.00 &     1.055        $\pm$     0.018 $\pm$ 0.016 \\
4.00 -- 4.50 &     1.077        $\pm$     0.041 $\pm$ 0.043 \\

\end{tabular}
\end{center}

\begin{center}
\begin{tabular}{cc}
\etamum & $R_{\ratiowmz}^{8/7}$ \\[0.1cm] \hline
2.00 -- 2.25 &     1.022        $\pm$     0.017 $\pm$ 0.018 \\
2.25 -- 2.50 &     0.969        $\pm$     0.015 $\pm$ 0.017 \\
2.50 -- 2.75 &     0.977        $\pm$     0.015 $\pm$ 0.013 \\
2.75 -- 3.00 &     0.925        $\pm$     0.015 $\pm$ 0.017 \\
3.00 -- 3.25 &     0.928        $\pm$     0.016 $\pm$ 0.016 \\
3.25 -- 3.50 &     0.906        $\pm$     0.017 $\pm$ 0.020 \\
3.50 -- 4.00 &     0.912        $\pm$     0.014 $\pm$ 0.014 \\
4.00 -- 4.50 &     0.922        $\pm$     0.026 $\pm$ 0.033 \\
\end{tabular}
\end{center}
\end{table}

\begin{table}[h]
\begin{center}
\caption{Cross-section for \Z boson production in bins of muon pseudorapidity at \sqs = 7~TeV. The first uncertainties are statistical, the second are systematic, the third are due to the knowledge of the \lhc beam energy and the fourth are due to the luminosity measurement. The last column lists the final-state radiation correction.}
\label{tab:csZETA7}
\begin{tabular}{cccc}
\etamu & $\csz(\etamup)$ [\pb] & $\ffsr^{\Z}$ \\ \hline
2.00 -- 2.25 &     12.69        $\pm$     0.13 $\pm$ 0.16 $\pm$ 0.16 $\pm$ 0.22           &$1.0290 \pm 0.0038$ \\
2.25 -- 2.50 &     12.31        $\pm$     0.13 $\pm$ 0.13 $\pm$ 0.15 $\pm$ 0.21           &$1.0246 \pm 0.0031$ \\
2.50 -- 2.75 &     11.27        $\pm$     0.12 $\pm$ 0.10 $\pm$ 0.14 $\pm$ 0.19           &$1.0237 \pm 0.0040$ \\
2.75 -- 3.00 &     10.16        $\pm$     0.11 $\pm$ 0.10 $\pm$ 0.13 $\pm$ 0.18           &$1.0242 \pm 0.0044$ \\
3.00 -- 3.25 &     \phantom{0}8.44        $\pm$     0.10 $\pm$ 0.08 $\pm$ 0.11 $\pm$ 0.15 &$1.0235 \pm 0.0033$ \\
3.25 -- 3.50 &     \phantom{0}7.15        $\pm$     0.10 $\pm$ 0.07 $\pm$ 0.09 $\pm$ 0.12 &$1.0258 \pm 0.0045$ \\
3.50 -- 4.00 &     \phantom{0}9.48        $\pm$     0.11 $\pm$ 0.08 $\pm$ 0.12 $\pm$ 0.16 &$1.0286 \pm 0.0032$ \\
4.00 -- 4.50 &     \phantom{0}4.49        $\pm$     0.08 $\pm$ 0.05 $\pm$ 0.06 $\pm$ 0.08 &$1.0386 \pm 0.0044$ \\
\end{tabular}

\vspace{0.5cm}

\begin{tabular}{cccc}
\etamu & $\csz(\etamum)$ [\pb] & $\ffsr^{\Z}$ \\ \hline
2.00 -- 2.25 &     11.93        $\pm$     0.13 $\pm$ 0.19 $\pm$ 0.15 $\pm$ 0.21           &$1.0294 \pm 0.0047$ \\
2.25 -- 2.50 &     11.61        $\pm$     0.12 $\pm$ 0.14 $\pm$ 0.15 $\pm$ 0.20           &$1.0251 \pm 0.0026$ \\
2.50 -- 2.75 &     11.12        $\pm$     0.12 $\pm$ 0.12 $\pm$ 0.14 $\pm$ 0.19           &$1.0245 \pm 0.0030$ \\
2.75 -- 3.00 &     \phantom{0}9.93        $\pm$     0.11 $\pm$ 0.10 $\pm$ 0.12 $\pm$ 0.17 &$1.0238 \pm 0.0031$ \\ 
3.00 -- 3.25 &     \phantom{0}8.91        $\pm$     0.11 $\pm$ 0.11 $\pm$ 0.11 $\pm$ 0.15 &$1.0236 \pm 0.0044$ \\ 
3.25 -- 3.50 &     \phantom{0}7.39        $\pm$     0.10 $\pm$ 0.08 $\pm$ 0.09 $\pm$ 0.13 &$1.0253 \pm 0.0048$ \\ 
3.50 -- 4.00 &     10.16        $\pm$     0.11 $\pm$ 0.11 $\pm$ 0.13 $\pm$ 0.18           &$1.0269 \pm 0.0047$ \\
4.00 -- 4.50 &     \phantom{0}4.95        $\pm$     0.08 $\pm$ 0.09 $\pm$ 0.06 $\pm$ 0.09 &$1.0376 \pm 0.0055$ \\ 
\end{tabular}
\end{center}
\end{table}

\begin{table}[h]
\begin{center}
\caption{(Top) \Wp and (bottom) \Wm to \Z cross-section ratios in bins of muon pseudorapidity at \sqs = 7~TeV. The first uncertainties are statistical, the second are systematic, the third are due to the knowledge of the \lhc beam energy and the fourth are due to the luminosity measurement.}
\label{tab:cswzr7}
\begin{tabular}{cc}
\etamup & \ratiowpz \\ \hline
2.00 -- 2.25 &     15.152        $\pm$     0.182 $\pm$ 0.231 $\pm$ 0.029 $\pm$ 0.001\\
2.25 -- 2.50 &     14.529        $\pm$     0.167 $\pm$ 0.230 $\pm$ 0.028 $\pm$ 0.001\\
2.50 -- 2.75 &     13.689        $\pm$     0.164 $\pm$ 0.176 $\pm$ 0.026 $\pm$ 0.001\\
2.75 -- 3.00 &     12.079        $\pm$     0.153 $\pm$ 0.153 $\pm$ 0.023 $\pm$ 0.001\\
3.00 -- 3.25 &     11.176        $\pm$     0.157 $\pm$ 0.151 $\pm$ 0.021 $\pm$ 0.001\\
3.25 -- 3.50 &     \phantom{0}8.623        $\pm$     0.134 $\pm$ 0.132 $\pm$ 0.016 $\pm$ 0.001\\
3.50 -- 4.00 &     \phantom{0}6.330        $\pm$     0.091 $\pm$ 0.076 $\pm$ 0.012 $\pm$ 0.001\\
4.00 -- 4.50 &     \phantom{0}3.198        $\pm$     0.101 $\pm$ 0.093 $\pm$ 0.006 $\pm$ 0.000\\
\end{tabular}
\end{center}

\begin{center}
\begin{tabular}{cc}
\etamum & \ratiowmz \\ \hline
2.00 -- 2.25 &     9.314        $\pm$     0.126 $\pm$ 0.162 $\pm$ 0.032 $\pm$ 0.001\\
2.25 -- 2.50 &     9.030        $\pm$     0.115 $\pm$ 0.151 $\pm$ 0.031 $\pm$ 0.001\\
2.50 -- 2.75 &     8.647        $\pm$     0.112 $\pm$ 0.112 $\pm$ 0.029 $\pm$ 0.001\\
2.75 -- 3.00 &     8.907        $\pm$     0.121 $\pm$ 0.150 $\pm$ 0.030 $\pm$ 0.001\\
3.00 -- 3.25 &     9.054        $\pm$     0.129 $\pm$ 0.156 $\pm$ 0.031 $\pm$ 0.001\\
3.25 -- 3.50 &     9.285        $\pm$     0.143 $\pm$ 0.202 $\pm$ 0.032 $\pm$ 0.001\\
3.50 -- 4.00 &     9.443        $\pm$     0.124 $\pm$ 0.126 $\pm$ 0.032 $\pm$ 0.001\\
4.00 -- 4.50 &     8.858        $\pm$     0.212 $\pm$ 0.243 $\pm$ 0.030 $\pm$ 0.001\\
\end{tabular}
\end{center}
\end{table}

\clearpage

\section{Correlation Matrices}
\label{sec:correlation}

\begin{sidewaystable}
\caption{Correlation coefficients between the differential \Wp and \Wm cross-sections in bins of muon $\eta$. The \lhc beam energy and luminosity uncertainties, which are fully correlated between cross-section measurements, are excluded.}
\label{tab:corr1}
\resizebox{24cm}{4cm}{
\begin{tabular}{l l | r r r r r r r r r r r r r r r r | r}
&\etamu & \multicolumn{2}{c}{\footnotesize{2--2.25}} & \multicolumn{2}{c}{\footnotesize{2.25--2.5}} & \multicolumn{2}{c}{\footnotesize{2.5--2.75}} & \multicolumn{2}{c}{\footnotesize{2.75--3}} & \multicolumn{2}{c}{\footnotesize{3--3.25}} & \multicolumn{2}{c}{\footnotesize{3.25--3.5}} & \multicolumn{2}{c}{\footnotesize{3.5--4}} & \multicolumn{2}{c}{\footnotesize{4--4.5}}  &  \\ \hline
&\multirow{2}{*}{\rotatebox[origin=c]{0}{\scriptsize{2--2.25}}}& 1\phantom{.00} & & & & & & & & & & & & & & & & \Wp\\
&&        0.67     &     1\phantom{.00}         & & & & & & & & & & & & & & & \Wm\\
&\multirow{2}{*}{\rotatebox[origin=c]{0}{\scriptsize{2.25--2.5}}}&        0.20    &    0.10    &      1\phantom{.00} & & & & & & & & & & & & & & \Wp\\        
&&        0.07    &    0.21    &    0.54     &     1\phantom{.00} & & & & & & & & & & & & & \Wm\\      
&\multirow{2}{*}{\rotatebox[origin=c]{0}{\scriptsize{2.5--2.75}}}&        0.13    &    0.24   &     0.12    &    0.23     &     1\phantom{.00} & & & & & & & & & & & & \Wp\\        
&&        0.05    &    0.18    &    0.03    &    0.22    &    0.64    &      1\phantom{.00} & & & & & & & & & & & \Wm\\      
&\multirow{2}{*}{\rotatebox[origin=c]{0}{\scriptsize{2.75--3}}}&        0.06    &    0.22    &    0.03    &    0.28    &    0.26    &    0.27     &     1\phantom{.00}  & & & & & & & & & & \Wp\\        
&&        0.04    &    0.21    &    0.00   &     0.25   &     0.25    &    0.31   &     0.70     &     1\phantom{.00} & & & & & & & & & \Wm\\        
&\multirow{2}{*}{\rotatebox[origin=c]{0}{\scriptsize{3--3.25}}}&        0.07    &    0.22    &    0.03    &    0.28    &    0.25    &    0.26   &     0.33    &    0.30    &      1\phantom{.00} & & & & & & & & \Wp\\        
&&        0.06    &    0.22    &    0.03    &    0.28   &     0.26   &     0.27   &     0.32    &    0.32   &     0.68     &     1\phantom{.00} & & & & & & & \Wm\\      
&\multirow{2}{*}{\rotatebox[origin=c]{0}{\scriptsize{3.25--3.5}}}&        0.03    &    0.23   &    $-$0.01    &    0.28    &    0.28    &    0.27    &    0.35   &     0.33    &    0.34   &     0.32     &     1\phantom{.00} & & & & & & \Wp\\        
&&        0.07   &     0.23    &    0.04    &    0.23    &    0.30    &    0.29    &    0.28    &    0.32   &     0.27   &     0.28   &     0.63     &     1\phantom{.00} & & & & & \Wm\\      
&\multirow{2}{*}{\rotatebox[origin=c]{0}{\scriptsize{3.5--4}}}&       $-$0.00  &      0.26   &    $-$0.06   &     0.33   &     0.31   &     0.32   &     0.41   &     0.39   &     0.40   &     0.38   &     0.45   &     0.36     &     1\phantom{.00}  & & & & \Wp\\      
&&        0.14   &    $-$0.06    &    0.20   &    $-$0.04   &    $-$0.13   &    $-$0.04   &    $-$0.08   &    $-$0.11   &    $-$0.07   &    $-$0.07   &    $-$0.17   &    $-$0.19   &    $-$0.04    &      1\phantom{.00} & & & \Wm\\      
&\multirow{2}{*}{\rotatebox[origin=c]{0}{\scriptsize{4--4.5}}}&       $-$0.07   &     0.14   &    $-$0.14    &    0.24    &    0.14   &     0.23   &     0.32   &     0.29   &     0.32   &     0.26    &    0.35   &     0.22   &     0.45   &    $-$0.15     &     1\phantom{.00} & & \Wp\\        
&&        0.12   &    $-$0.09    &    0.17    &   $-$0.11   &    $-$0.18   &    $-$0.14   &    $-$0.17   &    $-$0.19   &    $-$0.15   &    $-$0.18   &    $-$0.23   &    $-$0.24   &    $-$0.31    &    0.48    &    0.05     &     1\phantom{.00} & \Wm\\ \hline
&&\Wp & \Wm & \Wp & \Wm & \Wp & \Wm & \Wp & \Wm & \Wp & \Wm & \Wp & \Wm & \Wp & \Wm & \Wp & \Wm & \\
\end{tabular}
}
\end{sidewaystable}

\begin{sidewaystable}
\caption{Correlation coefficients between the differential \Z cross-section in bins of \rapz. The \lhc beam energy and luminosity uncertainties, which are fully correlated between cross-section measurements, are excluded.}
\label{tab:corr2}
\resizebox{24cm}{4cm}{
\begin{tabular}{c | c c c c c c c c c c c c c c c c c c}
\rapz & \scriptsize{2--2.125} & \scriptsize{2.125--2.25} & \scriptsize{2.25--2.375} & \scriptsize{2.375--2.5} & \scriptsize{2.5--2.625} &
\scriptsize{2.625--2.75} &
\scriptsize{2.75--2.875} & \scriptsize{2.875--3} & \scriptsize{3--3.125} &
\scriptsize{3.125--3.25} & \scriptsize{3.25--3.375} & \scriptsize{3.375--3.5} & \scriptsize{3.5--3.625} &
\scriptsize{3.625--3.75} &
\scriptsize{3.75--3.875} & \scriptsize{3.875--4} & \scriptsize{4--4.25}  & \scriptsize{4.25--4.5} \\ \hline
\scriptsize{2--2.125} &          1\phantom{.00}        & & & & & & & & & & & & & & & & & \\
\scriptsize{2.125--2.25} &        0.19   &       1\phantom{.00}        & & & & & & & & & & & & & & & &  \\
\scriptsize{2.25--2.375} &        0.17   &     0.27    &      1\phantom{.00}        & & & & & & & & & & & & & & &  \\
\scriptsize{2.375--2.5} &         0.16   &     0.26    &    0.28   &       1\phantom{.00}        &  & & & & & & & & & & & & & \\
\scriptsize{2.5--2.625} &        0.16    &    0.25     &   0.28    &    0.29   &       1\phantom{.00}         & & & & & & & & & & & & & \\
\scriptsize{2.625--2.75} &        0.15   &     0.24    &    0.27   &     0.29  &      0.30   &       1\phantom{.00}         & & & & & & & & & & & & \\
\scriptsize{2.75--2.875} &        0.14   &     0.23    &    0.26   &     0.28  &      0.29   &     0.30  &        1\phantom{.00}         &   & & & & & & & & & & \\
\scriptsize{2.875--3} &        0.14      &  0.21       & 0.25      &  0.27     &   0.29      &  0.30     &   0.30     &     1\phantom{.00}        & & & & & & & & & &  \\
\scriptsize{3--3.125} &        0.13      &  0.20       & 0.23      &  0.25     &   0.27      &  0.28     &   0.29     &   0.29      &    1\phantom{.00}        &  & & & & & & & & \\
\scriptsize{3.125--3.25} &        0.11   &     0.17    &    0.20   &     0.23  &      0.25   &     0.26  &      0.27  &      0.28   &     0.27  &        1\phantom{.00}        &   & & & & & & & \\
\scriptsize{3.25--3.375} &        0.09   &     0.14    &    0.16   &     0.18  &      0.20   &     0.22  &      0.22  &      0.23   &     0.23  &      0.23  &        1\phantom{.00}         &  & & & & & & \\
\scriptsize{3.375--3.5} &        0.08    &    0.12     &   0.15    &    0.17   &     0.19    &    0.20   &     0.21   &     0.22    &    0.22   &     0.22   &     0.20   &       1\phantom{.00}       &  & & & & & \\
\scriptsize{3.5--3.625} &        0.07    &    0.10     &   0.12    &    0.14   &     0.16    &    0.17   &     0.18   &     0.19    &    0.19   &     0.20   &     0.18   &     0.19   &       1\phantom{.00}        & & & & &  \\
\scriptsize{3.625--3.75} &        0.06   &     0.08    &    0.10   &     0.11  &      0.13   &     0.14  &      0.15  &      0.16   &     0.16  &      0.17  &      0.16  &      0.16  &      0.15  &        1\phantom{.00}        &  & & & \\
\scriptsize{3.75--3.875} &        0.05   &     0.07    &    0.08   &     0.09  &      0.10   &     0.11  &      0.11  &      0.12   &     0.13  &      0.13  &      0.12  &      0.13  &      0.12  &      0.11  &        1\phantom{.00}      &  & & \\
\scriptsize{3.875--4} &        0.03      &  0.05       & 0.06      &  0.06     &   0.07      &  0.08     &   0.08     &   0.09      &  0.09     &   0.10     &   0.09     &   0.10     &   0.09     &   0.08     &   0.07    &      1\phantom{.00}   & & \\    
\scriptsize{4--4.25} &        0.03       & 0.04        &0.04       & 0.05      &  0.05       & 0.06      &  0.06      &  0.06       & 0.07      &  0.07      &  0.07      &  0.08      &  0.07      &  0.07      &  0.06     &   0.05   &1\phantom{.00}  & \\  
\scriptsize{4.25--4.5} &        0.01     &   0.01      &  0.01     &   0.01    &    0.01     &   0.01    &    0.01     &   0.01     &   0.01    &    0.01    &   0.01     &   0.02     &   0.02     &   0.01     &   0.01    &    0.01   & 0.01  & 1\phantom{.00} \\
\end{tabular}
}
\end{sidewaystable}

\begin{sidewaystable}
\caption{Correlation coefficients between the differential \Z cross-section in bins of \ptz. The \lhc beam energy and luminosity uncertainties, which are fully correlated between cross-section measurements, are excluded.}
\label{tab:corr3}
\resizebox{24cm}{4cm}{
\begin{tabular}{c | c c c c c c c c c c c c c c }
\ptz [GeV/c]& \scriptsize{0.0--2.2} & \scriptsize{2.2--3.4} & \scriptsize{3.4--4.6} & \scriptsize{4.6--5.8} & \scriptsize{5.8--7.2} & \scriptsize{7.2--8.7}
& \scriptsize{8.7--10.5} & \scriptsize{10.5--12.8} &
\scriptsize{12.8--15.4} &
\scriptsize{15.4--19} & \scriptsize{19--24.5} & \scriptsize{24.5--34} & \scriptsize{34--63} &
\scriptsize{63--270} \\ \hline
\scriptsize{0.0--2.2} &          1\phantom{.00}       & & & & & & & & & & & & & \\
\scriptsize{2.2--3.4} &        0.06&          1\phantom{.00}        & & & & & & & & & & & & \\
\scriptsize{3.4--4.6} &        0.08&        0.16&          1\phantom{.00}         & & & & & & & & & & & \\
\scriptsize{4.6--5.8} &        0.13&        0.09&        0.20&          1\phantom{.00}        & & & & & & & & & & \\
\scriptsize{5.8--7.2} &        0.15&        0.16&        0.12&        0.19&          1\phantom{.00}        & & & & & & & & & \\
\scriptsize{7.2--8.7} &        0.14&        0.16&        0.18&        0.11&        0.17&          1\phantom{.00}        & & & & & & & & \\
\scriptsize{8.7--10.5} &        0.14&        0.16&        0.20&        0.19&        0.14&        0.15&          1\phantom{.00}        & & & & & & & \\
\scriptsize{10.5--12.8} &        0.14&        0.16&        0.19&        0.19&        0.21&        0.14&        0.12&          1\phantom{.00}        & & & & & & \\
\scriptsize{12.8--15.4} &        0.15&        0.17&        0.20&        0.19&        0.22&        0.20&        0.17&        0.11&          1\phantom{.00}        & & & & & \\
\scriptsize{15.4--19} &        0.14&        0.16&        0.20&        0.19&        0.21&        0.19&        0.21&        0.18&        0.11&          1\phantom{.00}       & & & & \\
\scriptsize{19--24.5} &        0.15&        0.17&        0.21&        0.20&        0.22&        0.20&        0.21&        0.21&        0.21&        0.13&          1\phantom{.00}    & & &\\    
\scriptsize{24.5--34} &        0.16&        0.18&        0.22&        0.21&        0.23&        0.21&        0.22&        0.22&        0.23&        0.22&        0.17&          1\phantom{.00} & &\\      
\scriptsize{34--63} &        0.16&        0.18&        0.22&        0.21&        0.23&        0.21&        0.22&        0.22&        0.23&        0.22&        0.23&        0.21&          1\phantom{.00} & \\      
\scriptsize{63--270} &        0.11&        0.12&        0.15&        0.14&        0.16&        0.15&        0.15&        0.15&        0.16&        0.15&        0.16&        0.17&        0.15&          1\phantom{.00} \\
\end{tabular}
}
\end{sidewaystable}

\begin{sidewaystable}
\caption{Correlation coefficients between the differential \Z cross-section in bins of \phist. The \lhc beam energy and luminosity uncertainties, which are fully correlated between cross-section measurements, are excluded.}
\label{tab:corr4}
\resizebox{24cm}{4cm}{
\begin{tabular}{c | c c c c c c c c c c c c c c c }
\phist & \scriptsize{0.00--0.01} & \scriptsize{0.01--0.02} & \scriptsize{0.02--0.03} & \scriptsize{0.03--0.05}
& \scriptsize{0.05--0.07} & \scriptsize{0.07--0.10} & \scriptsize{0.10--0.15} & \scriptsize{0.15--0.20} &\scriptsize{0.20--0.30}
& \scriptsize{0.30--0.40} & \scriptsize{0.40--0.60} & \scriptsize{0.60--0.80} & \scriptsize{0.80--1.20} & \scriptsize{1.20--2.00} & \scriptsize{2.00--4.00} \\ \hline
\scriptsize{0.00--0.01} &          1\phantom{.00}         & & & & & & & & & & & & & & \\
\scriptsize{0.01--0.02} &        0.50&          1\phantom{.00}        & & & & & & & & & & & & & \\
\scriptsize{0.02--0.03} &        0.42&        0.39&          1\phantom{.00}        & & & & & & & & & & & & \\
\scriptsize{0.03--0.05} &        0.57&        0.55&        0.44&          1\phantom{.00}        & & & & & & & & & & & \\
\scriptsize{0.05--0.07} &        0.52&        0.50&        0.41&        0.55&          1\phantom{.00}        & & & & & & & & & & \\
\scriptsize{0.07--0.10} &        0.44&        0.42&        0.34&        0.47&        0.42&          1\phantom{.00}        & & & & & & & & & \\
\scriptsize{0.10--0.15} &        0.50&        0.48&        0.40&        0.54&        0.49&        0.41&          1\phantom{.00}        & & & & & & & & \\
\scriptsize{0.15--0.20} &        0.49&        0.47&        0.38&        0.52&        0.48&        0.40&        0.46&          1\phantom{.00}        & & & & & & &\\
\scriptsize{0.20--0.30} &        0.47&        0.45&        0.37&        0.50&        0.45&        0.39&        0.44&        0.43&          1\phantom{.00}        & & & & & &\\
\scriptsize{0.30--0.40} &        0.31&        0.30&        0.24&        0.33&        0.30&        0.26&        0.29&        0.29&        0.27&          1\phantom{.00}        & & & & &\\
\scriptsize{0.40--0.60} &        0.31&        0.29&        0.24&        0.33&        0.30&        0.25&        0.29&        0.28&        0.27&        0.18&          1\phantom{.00}      & & & &\\  
\scriptsize{0.60--0.80} &        0.21&        0.20&        0.16&        0.23&        0.20&        0.17&        0.20&        0.19&        0.19&        0.12&        0.12&          1\phantom{.00}  & & &\\    
\scriptsize{0.80--1.20} &        0.13&        0.13&        0.10&        0.14&        0.13&        0.11&        0.13&        0.12&        0.12&        0.08&        0.08&        0.05&          1\phantom{.00}  & &\\    
\scriptsize{1.20--2.00} &        0.07&        0.07&        0.06&        0.08&        0.07&        0.06&        0.07&        0.07&        0.06&        0.04&        0.04&        0.03&        0.02&          1\phantom{.00} &\\    
\scriptsize{2.00--4.00} &        0.02&        0.02&        0.02&        0.02&        0.02&        0.02&        0.02&        0.02&        0.02&        0.01&        0.01&        0.01&        0.01&        0.00&          1\phantom{.00} \\
\end{tabular}
}
\end{sidewaystable}

\begin{sidewaystable}
\caption{Correlation coefficients between differential \W and \Z cross-sections in bins of muon $\eta$ and \rapz, respectively. The \lhc beam energy and luminosity uncertainties, which are fully correlated between cross-section measurements, are excluded.}
\label{tab:corr5}
\resizebox{24cm}{4cm}{
\begin{tabular}{l l | c c c c c c c c c c c c c c c c c c | r}
 && \multicolumn{18}{c}{\rapz} & \\
 && \scriptsize{2--2.125} & \scriptsize{2.125--2.25} & \scriptsize{2.25--2.375} & \scriptsize{2.375--2.5} & \scriptsize{2.5--2.625} & \scriptsize{2.625--2.75} &\scriptsize{2.75--2.875} & \scriptsize{2.875--3} & \scriptsize{3--3.125} & \scriptsize{3.125--3.25} & \scriptsize{3.25--3.375} & \scriptsize{3.375--3.5} & \scriptsize{3.5--3.625} & \scriptsize{3.625--3.75} & \scriptsize{3.75--3.875} & \scriptsize{3.875--4} & \scriptsize{4--4.25}  & \scriptsize{4.25--4.5} &  \\ \hline
\multirow{16}{*}{\rotatebox[origin=c]{0}{\etamu}}&\multirow{2}{*}{\rotatebox[origin=c]{0}{\scriptsize{2--2.25}}}&       0.23 &        0.30 &       0.28 &        0.27 &       0.26 &      0.25 &        0.24 &        0.23 &       0.21 &       0.18 &       0.15 &       0.13 &       0.11 &      0.10 &       0.07 &      0.05 &      0.04 &     0.01 & \Wp\\
&&       0.21 &        0.28 &       0.26 &        0.25 &       0.24 &        0.24 &        0.22 &        0.21 &       0.20 &       0.17 &       0.14 &       0.12 &       0.11 &      0.09 &       0.07 &      0.05 &      0.04 &     0.01 & \Wm \\
&\multirow{2}{*}{\rotatebox[origin=c]{0}{\scriptsize{2.25--2.5}}}&       0.05 &        0.15 &       0.21 &        0.20 &       0.20 &        0.20 &        0.20 &        0.19 &       0.18 &       0.16 &       0.14 &       0.12 &       0.10 &      0.08 &       0.06 &      0.04 &      0.03 &     0.01 & \Wp\\
&&       0.04 &        0.14 &       0.19 &        0.18 &       0.18 &        0.18 &        0.18 &        0.17 &       0.16 &       0.15 &       0.12 &       0.11 &       0.09 &      0.07 &       0.05 &      0.04 &      0.03 &     0.00 & \Wm \\
&\multirow{2}{*}{\rotatebox[origin=c]{0}{\scriptsize{2.5--2.75}}}&       0.04 &        0.07 &       0.12 &        0.15 &       0.16 &        0.17 &        0.17 &        0.17 &       0.16 &       0.15 &       0.13 &       0.13 &       0.11 &      0.08 &       0.06 &      0.05 &      0.03 &     0.01 & \Wp\\
&&       0.04 &        0.07 &       0.11 &        0.14 &       0.15 &        0.15 &        0.15 &        0.15 &       0.15 &       0.14 &       0.12 &       0.12 &       0.10 &      0.08 &       0.06 &      0.04 &      0.03 &     0.01 & \Wm \\
&\multirow{2}{*}{\rotatebox[origin=c]{0}{\scriptsize{2.75--3}}}&       0.05 &        0.08 &       0.10 &        0.13 &       0.16 &        0.17 &        0.17 &        0.17 &       0.16 &       0.16 &       0.14 &       0.13 &       0.12 &      0.09 &       0.07 &      0.05 &      0.03 &     0.01 & \Wp\\
&&       0.05 &        0.07 &       0.09 &        0.12 &       0.14 &        0.15 &        0.15 &        0.16 &       0.15 &       0.14 &       0.12 &       0.12 &       0.11 &      0.08 &     0.06 &      0.04 &      0.03 &     0.01 & \Wm \\
&\multirow{2}{*}{\rotatebox[origin=c]{0}{\scriptsize{3--3.25}}}&       0.06 &        0.08 &       0.10 &        0.11 &       0.14 &        0.16 &        0.17 &        0.17 &       0.16 &       0.16 &       0.14 &       0.14 &       0.12 &      0.10 &       0.07 &      0.05 &      0.03 &     0.01 & \Wp\\
&&       0.05 &        0.08 &       0.09 &        0.10 &       0.13 &        0.15 &        0.15 &        0.16 &       0.15 &       0.15 &       0.13 &       0.13 &       0.11 &      0.09 &     0.07 &      0.05 &      0.03 &     0.01 & \Wm \\
&\multirow{2}{*}{\rotatebox[origin=c]{0}{\scriptsize{3.25--3.5}}}&       0.04 &        0.06 &       0.07 &        0.09 &       0.10 &        0.11 &        0.12 &        0.13 &       0.13 &     0.12 &       0.11 &       0.11 &       0.09 &      0.08 &       0.06 &      0.04 &      0.03 &     0.00 & \Wp\\
&&       0.04 &        0.06 &       0.08 &        0.09 &       0.10 &        0.12 &        0.13 &        0.13 &       0.13 &       0.13 &       0.11 &       0.11 &       0.10 &      0.08 &     0.06 &      0.04 &      0.03 &     0.00 & \Wm \\
&\multirow{2}{*}{\rotatebox[origin=c]{0}{\scriptsize{3.5--4}}}&       0.04 &        0.06 &       0.07 &        0.08 &       0.09 &        0.10 &        0.11 &        0.12 &       0.12 &       0.12 &       0.11 &       0.11 &       0.10 &      0.09 &       0.07 &      0.05 &      0.04 &     0.00 & \Wp\\
&&       0.04 &        0.06 &       0.08 &        0.09 &       0.10 &        0.11 &        0.12 &        0.13 &       0.14 &       0.14 &       0.12 &       0.12 &       0.11 &      0.10 &    0.08 &      0.06 &      0.04 &     0.01 & \Wm \\
&\multirow{2}{*}{\rotatebox[origin=c]{0}{\scriptsize{4--4.5}}}&       0.02 &        0.03 &       0.04 &        0.04 &       0.04 &        0.04 &        0.05 &        0.05 &       0.06 &       0.06 &       0.06 &       0.07 &       0.06 &      0.06 &       0.05 &      0.04 &      0.04 &     0.01 & \Wp\\
&&       0.03 &        0.04 &       0.04 &        0.05 &       0.05 &        0.06 &        0.06 &        0.06 &       0.07 &       0.07 &       0.07 &       0.08 &       0.08 &      0.07 &    0.06 &      0.05 &      0.04 &     0.01 & \Wm \\
\end{tabular}
}
\end{sidewaystable}

\begin{table}[h]
\begin{center}
\caption{Correlation coefficients between the differential \Z cross-sections in bins of (top) \etamup and (bottom) \etamum. The \lhc beam energy and luminosity uncertainties, which are fully correlated between cross-section measurements, are excluded.}
\label{tab:corr6}
\begin{tabular}{c | c c c c c c c c }
\etamup& \scriptsize{2.00--2.25} & \scriptsize{2.25--2.50} & \scriptsize{2.50--2.75} & \scriptsize{2.75--3.00}
& \scriptsize{3.00--3.25} & \scriptsize{3.25--3.50} & \scriptsize{3.50--4.00} & \scriptsize{4.00--4.50} \\ \hline
\scriptsize{2.00--2.25}&          1\phantom{.00}        & & & & & & &  \\
\scriptsize{2.25--2.50}&        0.31&          1\phantom{.00}        & & & & & &  \\
\scriptsize{2.50--2.75}&        0.30&        0.27&          1\phantom{.00}       & & & & &  \\
\scriptsize{2.75--3.00}&        0.31&        0.28&        0.26&          1\phantom{.00}    & & & &  \\    
\scriptsize{3.00--3.25}&        0.32&        0.28&        0.26&        0.27&          1\phantom{.00}  & & &  \\      
\scriptsize{3.25--3.50}&        0.27&        0.25&        0.23&        0.23&        0.23&          1\phantom{.00}  & &  \\        
\scriptsize{3.50--4.00}&        0.33&        0.30&        0.28&        0.28&        0.28&        0.25&          1\phantom{.00}  &  \\      
\scriptsize{4.00--4.50}&        0.28&        0.25&        0.23&        0.24&        0.23&        0.20&        0.23&          1\phantom{.00} \\
\end{tabular}
\end{center}

\begin{center}
\begin{tabular}{c | c c c c c c c c }
\etamum& \scriptsize{2.00--2.25} & \scriptsize{2.25--2.50} & \scriptsize{2.50--2.75} & \scriptsize{2.75--3.00}
& \scriptsize{3.00--3.25} & \scriptsize{3.25--3.50} & \scriptsize{3.50--4.00} & \scriptsize{4.00--4.50} \\ \hline
\scriptsize{2.00--2.25}&          1\phantom{.00}        & & & & & & &  \\
\scriptsize{2.25--2.50}&        0.29&          1\phantom{.00}        & & & & & &  \\
\scriptsize{2.50--2.75}&        0.30&        0.29&          1\phantom{.00}        & & & & &  \\
\scriptsize{2.75--3.00}&        0.30&        0.28&        0.28&          1\phantom{.00}        & & & &  \\
\scriptsize{3.00--3.25}&        0.30&        0.27&        0.27&        0.27&          1\phantom{.00}      & & &  \\  
\scriptsize{3.25--3.50}&        0.27&        0.25&        0.25&        0.24&        0.24&          1\phantom{.00}  & &  \\      
\scriptsize{3.50--4.00}&        0.31&        0.29&        0.29&        0.28&        0.28&        0.25&          1\phantom{.00}  &  \\      
\scriptsize{4.00--4.50}&        0.28&        0.25&        0.25&        0.24&        0.24&        0.21&        0.24&          1\phantom{.00} \\
\end{tabular}
\end{center}
\end{table}

\clearpage

\begin{table}[h]
\begin{center}
\caption{Correlation coefficients between the differential \W and \Z cross-sections in bins of (top) \etamup and (bottom) \etamum. The \lhc beam energy and luminosity 
uncertainties, which are fully correlated between cross-section measurements, are excluded.}
\label{tab:corr7}
\begin{tabular}{c c | c c c c c c c c }
& & \multicolumn{8}{c}{\Z} \\
&\etamup& \scriptsize{2.00--2.25} & \scriptsize{2.25--2.50} & \scriptsize{2.50--2.75} & \scriptsize{2.75--3.00}
& \scriptsize{3.00--3.25} & \scriptsize{3.25--3.50} & \scriptsize{3.50--4.00} & \scriptsize{4.00--4.50} \\ \hline
\multirow{8}{*}{\rotatebox[origin=c]{0}{\W}}&\scriptsize{2.00--2.25}&         0.48&        0.19&        0.19&        0.20&        0.20&        0.18&        0.22&        0.18\\
&\scriptsize{2.25--2.50}&        0.15&        0.38&        0.16&        0.16&        0.16&        0.14&        0.17&        0.14\\
&\scriptsize{2.50--2.75}&        0.14&        0.14&        0.26&        0.14&        0.14&        0.13&        0.16&        0.12\\
&\scriptsize{2.75--3.00}&        0.14&        0.14&        0.14&        0.26&        0.15&        0.13&        0.16&        0.12\\
&\scriptsize{3.00--3.25}&        0.15&        0.14&        0.14&        0.15&        0.25&        0.13&        0.15&        0.12\\
&\scriptsize{3.25--3.50}&        0.11&        0.11&        0.11&        0.11&        0.11&        0.17&        0.12&        0.09\\
&\scriptsize{3.50--4.00}&        0.10&        0.10&        0.11&        0.11&        0.11&        0.10&        0.18&        0.08\\
&\scriptsize{4.00--4.50}&        0.06&        0.06&        0.06&        0.06&        0.06&        0.05&        0.06&        0.11\\

\end{tabular}
\end{center}

\begin{center}
\begin{tabular}{c c | c c c c c c c c }
& & \multicolumn{8}{c}{\Z} \\
&\etamum& \scriptsize{2.00--2.25} & \scriptsize{2.25--2.50} & \scriptsize{2.50--2.75} & \scriptsize{2.75--3.00}
& \scriptsize{3.00--3.25} & \scriptsize{3.25--3.50} & \scriptsize{3.50--4.00} & \scriptsize{4.00--4.50} \\ \hline
\multirow{8}{*}{\rotatebox[origin=c]{0}{\W}}&\scriptsize{2.00--2.25}&        0.43&        0.18&        0.19&        0.19&        0.19&        0.18&        0.21&        0.18\\
&\scriptsize{2.25--2.50}&        0.13&        0.35&        0.15&        0.15&        0.14&        0.14&        0.16&        0.13\\
&\scriptsize{2.50--2.75}&        0.12&        0.13&        0.25&        0.13&        0.13&        0.12&        0.14&        0.12\\
&\scriptsize{2.75--3.00}&        0.12&        0.13&        0.14&        0.24&        0.13&        0.12&        0.14&        0.12\\
&\scriptsize{3.00--3.25}&        0.13&        0.13&        0.14&        0.14&        0.23&        0.13&        0.15&        0.12\\
&\scriptsize{3.25--3.50}&        0.11&        0.11&        0.12&        0.12&        0.12&        0.18&        0.12&        0.10\\
&\scriptsize{3.50--4.00}&        0.11&        0.12&        0.12&        0.12&        0.12&        0.11&        0.20&        0.10\\
&\scriptsize{4.00--4.50}&        0.07&        0.06&        0.07&        0.07&        0.07&        0.06&        0.07&        0.13\\

\end{tabular}
\end{center}
\end{table}

\addcontentsline{toc}{section}{References}
\setboolean{inbibliography}{true}
\bibliographystyle{LHCb}
\bibliography{main,LHCb-PAPER,LHCb-CONF,LHCb-DP,LHCb-TDR}

\newpage

\newpage
\centerline{\large\bf LHCb collaboration}
\begin{flushleft}
\small
R.~Aaij$^{39}$, 
C.~Abell\'{a}n~Beteta$^{41}$, 
B.~Adeva$^{38}$, 
M.~Adinolfi$^{47}$, 
A.~Affolder$^{53}$, 
Z.~Ajaltouni$^{5}$, 
S.~Akar$^{6}$, 
J.~Albrecht$^{10}$, 
F.~Alessio$^{39}$, 
M.~Alexander$^{52}$
% ...

{\footnotesize \it
% ...
$ ^{5}$Clermont Universit\'{e}, Universit\'{e} Blaise Pascal, CNRS/IN2P3, LPC, Clermont-Ferrand, France\\
$ ^{6}$CPPM, Aix-Marseille Universit\'{e}, CNRS/IN2P3, Marseille, France\\
% ...
$ ^{10}$Fakult\"{a}t Physik, Technische Universit\"{a}t Dortmund, Dortmund, Germany\\
% ...
$ ^{38}$Universidad de Santiago de Compostela, Santiago de Compostela, Spain\\
$ ^{39}$European Organization for Nuclear Research (CERN), Geneva, Switzerland\\
% ...
$ ^{41}$Physik-Institut, Universit\"{a}t Z\"{u}rich, Z\"{u}rich, Switzerland\\
% ...
$ ^{47}$H.H. Wills Physics Laboratory, University of Bristol, Bristol, United Kingdom\\
% ...
$ ^{52}$School of Physics and Astronomy, University of Glasgow, Glasgow, United Kingdom\\
$ ^{53}$Oliver Lodge Laboratory, University of Liverpool, Liverpool, United Kingdom\\
% ...
}
\end{flushleft}

\end{document}